\documentclass[aps,twocolumn,showkeys,showpacs,preprintnumbers,prd,superscriptaddress,nofootinbib,colorlinks=true,linkcolor=blue,urlcolor=blue,citecolor=blue,anchorcolor=blue]{revtex4-2}

\usepackage{comment}
\usepackage{graphicx}
\usepackage{epsf}
\usepackage{bm}
\usepackage{amsmath}
\usepackage{amsfonts}
\usepackage{amssymb}
\usepackage{epstopdf}
\usepackage{natbib}
\usepackage{color}
\usepackage{array}
\usepackage{algorithmicx}
\usepackage{algorithm}
\usepackage[noend]{algpseudocode}

\usepackage[dvipsnames]{xcolor}
\usepackage{verbatim}
\usepackage{soul}
\usepackage{physics}

\usepackage{titlesec}
\usepackage[T1]{fontenc}
\usepackage{xcolor}
\usepackage{verbatim}
\usepackage[utf8]{inputenc} 
\usepackage{amsmath,amssymb} 
\usepackage{graphicx} 
\usepackage{bm}
\usepackage{appendix}
\usepackage{subcaption}

\usepackage{orcidlink}

\usepackage[capitalize]{cleveref}
\usepackage{subcaption}
\hypersetup{
    linktoc=all,
    colorlinks=true,
    linkcolor=blue,
    citecolor=red,
    urlcolor=magenta
}

\usepackage{bm}
\usepackage{microtype}

\usepackage{amsmath}
\usepackage[capitalize]{cleveref}
\usepackage[normalem]{ulem}
\usepackage{enumitem}

\usepackage{booktabs}
\usepackage{multirow}

\usepackage{lipsum}

\begin{document}

\title{Cosmological Parameter Estimation using Particle Swarm Optimization}

\author{Daniel Morales Hernández}
\affiliation{Instituto de Física, Universidad Nacional Autónoma de México, Ciudad de México, México}

\author{Gabriela Garcia-Arroyo\,\orcidlink{0000-0002-0599-7036}}
\email{arroyo@icf.unam.mx}
\affiliation{Instituto de Ciencias F\'isicas, Universidad Nacional Aut\'onoma de M\'exico, Cuernavaca, 
Morelos, 62210, M\'exico}

\author{J. Alberto Vazquez\,\orcidlink{0000-0002-7401-0864}}
\email{javazquez@icf.unam.mx}
\affiliation{Instituto de Ciencias F\'isicas, Universidad Nacional Aut\'onoma de M\'exico, Cuernavaca, 
Morelos, 62210, M\'exico}

\begin{abstract}

The quest for a theoretical framework and ingredients that capture our current understanding of the cosmos has motivated the design of a large number of highly informative experiments, generating an abundant flow of data. Faced with this amount of data and the need for thorough analysis, the main aim of this work is to present and assess the Particle Swarm Optimization (PSO) algorithm as a complementary tool to conventional cosmological data analysis techniques. PSO is one of the most representative bio-inspired algorithms, offering strong robustness for high-dimensional or complex problems, while remaining relatively simple to implement and requiring only a few  hyperparameters. 
In this study, we employ two standard variants of the canonical PSO algorithm—global best and local best—to investigate dark energy models using Type Ia Supernovae and Baryon Acoustic Oscillation measurements, focusing in particular on the DESI and DESI+Union3 datasets. 
Our findings demonstrate that PSO effectively recovers the best-fit parameters from observational data and show that, under suitable conditions, PSO can achieve results comparable to those of traditional MCMC techniques but in a significantly reduced computation time. 
Moreover, the solutions obtained with PSO can be used as high-quality initial conditions for MCMC analyzes, thereby accelerating their convergence.

\end{abstract}


\maketitle

\section{Introduction}

Current measurements of the Universe offer compelling evidence that it is not only expanding but also doing so at an accelerating rate. The evidence for the expansion started with Vesto Slipher’s redshift data~\cite{slipher1913radial}, followed by Edwin Hubble’s distance–velocity relation for galaxies~\cite{hubble1929relation}, whereas the accelerated expansion was established later by Type Ia supernovae (SNIa) observations~\cite{riess1998observational, perlmutter1999measurements}.
 These findings, combined with General Relativity, established the foundations of modern cosmology and led to the establishment of the $\Lambda$CDM model, often referred to as the concordance model due to its consistency with a wide range of observational data, including the Cosmic Microwave Background (CMB) \cite{aghanim2020planck}, Baryon Acoustic Oscillations (BAO) \cite{eisenstein2005detection}, Large-Scale Structure (LSS) \cite{SDSS:2003eyi}, and SNIa data~\cite{suzuki2012hubble}, among many others.

Despite its simplicity and remarkable success, the $\Lambda$CDM model still faces several challenges \cite{Bull:2015stt}. These include small-scale problems, such as the Core-Cusp problem and the Too Big to Fail problem \cite{vazquez2008materia}. Moreover, the lack of compelling results in the direct detection of dark matter continues to raise questions about its nature. 
The model is further affected by the cosmological constant problem, $\Lambda$, namely the significant discrepancy between the observed value and its predictions by quantum field theory \cite{weinberg1989cosmological, shi2012comprehensive}. Another prominent issue is the so-called Hubble tension, referring to the mismatch between local determinations of the Hubble constant $H_0$~\cite{Riess:2021jrx} and the value inferred from CMB observations, assuming the $\Lambda$CDM framework \cite{verde2014expansion}. 
Moreover, recent BAO measurements from the Dark Energy Spectroscopic Instrument (DESI), when analyzed jointly with CMB and SNIa data, put the $\Lambda$CDM model under strain at more than $3\sigma$ significance and instead point towards a dynamical dark energy component characterized by at least two additional parameters \cite{DESI:2024mwx, DESI:2025zgx}.
These discrepancies motivate exploring extensions for the $\Lambda$CDM model by relaxing one or more of its fundamental assumptions~\cite{CosmoVerseNetwork:2025alb, arun2017dark}. In particular, dark energy (DE) scenarios featuring a time-dependent equation of state (EoS) have been widely investigated as alternatives to the cosmological constant~\cite{copeland2006dynamics, Linder:2002et, Chevallier:2000qy, Escamilla:2024fzq}, allowing for a more versatile characterization of cosmic acceleration.
However, such extensions come with extra free parameters, which increase the model’s complexity and, consequently, make conventional inference methods more computationally expensive.

A central challenge in contemporary cosmology is to infer the parameter values that best reproduce the observational data for a given theoretical model, a problem usually tackled with Markov Chain Monte Carlo (MCMC) techniques (For 
 an extensive review of MCMC algorithms in Cosmology, see~\cite{lewis2002cosmological,  trotta2017bayesian, gilks1995markov, padilla2021cosmological}).
Despite their widespread use, these methods can become computationally expensive when the parameter space is high-dimensional, when the parameters are strongly correlated, or when the likelihood surface contains several local maxima or even discontinuities \cite{Akarsu:2019hmw, Vazquez:2020ani}.
Such drawbacks motivate the exploration of alternative optimization strategies to accelerate convergence, increase efficiency, and improve the sampling of complex, high-dimensional parameter spaces~\cite{Gomez-Vargas:2024izm, Medel-Esquivel:2023nov, Gomez-Vargas:2022bsm}.
\\

In this work, we employ the Particle Swarm Optimization (PSO) algorithm, originally proposed by James Kennedy and Russel C. Eberhart in 1995~\cite{kennedy1995particle} {(inspired by computer simulations of social models investigated by Reynolds in 1987 \cite{reynolds1987flocks}),
to perform the parameter inference from a collection of cosmological models.
The PSO method was initially motivated by the collective behavior observed in social organisms, such as flocks of birds searching for food or shelter. 
PSO is part of the family of \emph{bioinspired algorithms}, a category that also includes genetic algorithms (GA) and evolutionary programming (EP), both of which fall under the broader group of Evolutionary Algorithms \cite{darwish2018bio}; in addition to the PSO, the Ant Colony Optimization Algorithm (ACO) and Artificial Bee Colony (ABC) are classified as Swarm Algorithms.
It is worth noting that numerous extensions and modifications of the original PSO, introduced by Eberhart and Kennedy, have been developed since then, such as Cooperative Particle Swarm Optimization (CPSO) \cite{van2000cooperative} and Adaptive Particle Swarm Optimization (APSO) \cite{zhan2009adaptive}, among many others \cite{freitas2020particle}. The primary distinction between the basic PSO and its variants lies in the number and type of hyper-parameters. While the original PSO relies mainly on random numbers to update particle positions and velocities, several variants introduce acceleration coefficients ($c_{1}$ and $c_{2}$) and an inertia weight ($w$), the latter added by Shi and Eberhart \cite{shi1998modified} to regulate the swarm’s exploration and exploitation capabilities \cite{shi2001particle,  engelbrecht2007computational}. These enhancements led to what is commonly known as the \emph{canonical PSO} \cite{sengupta2018particle}, which we will describe in the following sections.
\\

The widespread adoption of PSO is largely due to its strong robustness in searching over high-dimensional spaces or dealing with complex optimization problems while remaining relatively simple to implement and requiring only a few hyperparameters. In particular, unlike genetic algorithms, PSO does not rely on explicit selection or mutation operators. By combining the intrinsic properties of PSO with the exploratory behavior, communication, and mutual influence of swarm particles, the algorithm can efficiently identify promising regions of the search space with high accuracy and in a significantly reduced runtime~\cite{engelbrecht2007computational}.  
Consequently, the PSO algorithm is especially well suited for cosmological models characterized by high-dimensional parameter spaces and computationally expensive likelihood functions, which is precisely the scenario anticipated in upcoming cosmological analyzes.  
PSO has already seen extensive use across multiple fields, including engineering and medicine \cite{poli2008analysis, abido2002optimal}, as well as astrophysics and cosmology \cite{prasad2012cosmological, skokos2005particle, wang2015first}. 
In particular, the authors in \cite{prasad2012cosmological} applied PSO to cosmological parameter estimation for the standard $\Lambda$CDM model using CMB data \cite{WMAP:2010qai}, while \cite{skokos2005particle} and \cite{wang2015first} employed it to trace periodic orbits in three-dimensional galactic potentials and for gravitational-wave data analysis, respectively.
It is important to emphasize that PSO is not aiming to replace MCMC methods, which remain standard tools in cosmology. MCMC algorithms are designed to sample the entire posterior distribution, whereas PSO focuses on efficiently locating the maximum of the likelihood function, similar to the role of genetic algorithms \cite{Medel-Esquivel:2023nov}.
Modern Bayesian inference frameworks already incorporate optimization routines as complementary tools to posterior sampling \cite{Zuntz:2014csq, Torrado:2020dgo}. Likewise, alternative population-based Bayesian approaches have also been developed \cite{Toni_2008, toni2010simulationbasedmodelselectiondynamical}   and successfully applied to test dark energy models \cite{Bernardo:2022pyz, Bernardo:2022ggl}. 
Nonetheless, the search trajectory generated by both approaches—across generations in GA or iterations in PSO—encodes useful information about the underlying posterior distribution.

The primary goal of this work is to introduce an implementation of a code capable of performing rapid and precise cosmological parameter estimation for a specific model and any chosen combination of datasets. As a proof of concept, we have restricted our study to background-level observations and leave the incorporation of linear perturbations into the analysis for future work.
We employ the two variants of the standard PSO algorithm — global best and local best — to study dark energy models using, i.e., Type Ia Supernovae and Baryon Acoustic Oscillation measurements, focusing in particular on the DESI and DESI+Union3 datasets. It is worth emphasizing that both the selected model and the dataset combinations can be easily adjusted within the code in order to explore a wide diversity of combinations.  
The full module generated in this analysis is publicly accessible as a modified component of the \href{https://github.com/ja-vazquez/SimpleMC/blob/master/simplemc/analyzers/PSO_optimizer.py}{SimpleMC} code \cite{BOSS:2014hhw}.  
Finally, using the best-fit parameters, we compute the Akaike and Bayesian information criteria to perform model selection, and we apply the Fisher matrix formalism to estimate confidence intervals and produce error plots, which we then compare against MCMC posterior distributions.
\\

The paper is organized as follows. Section~\ref{Section-2} presents the theoretical foundation and introduces the principal equations that describe the dark energy models examined in this work. 
Building on this basis, Section~\ref{Sec:paraeter_estimation} describes the inference methodology, emphasizing the use of PSO for parameter estimation and defining the criteria for model comparison, along with the observational data sets employed.
Subsequently, Section~\ref{Section_PSO} provides a detailed account of the PSO algorithm, including particle dynamics, the choice of hyperparameters, and the distinctions between the Global Best and Local Best schemes.
The main target of the paper is addressed in Section~\ref{Seccion-7}, where we implement the Global Best PSO for cosmological parameter estimation and present the resulting parameter values and model comparisons. Finally, Section~\ref{Sec:conclusions} summarizes the main results and discusses the significance of PSO in cosmological studies.

\section{Cosmological Framework}\label{Section-2}

The standard cosmological model presumes that, on sufficiently large scales, the Universe is both homogeneous and isotropic. Under this assumption, spacetime is described by the Friedmann–Lemaître–Robertson–Walker (FLRW) metric:
\begin{equation}
   ds^{2}=dt^{2}-a^{2}(t) \left(\frac{dr^{2}}{1-kr^{2}}+ r^{2}d\theta^{2} + r^{2}\sin^{2}(\theta)d\phi^{2} \right)\,,
   \label{metrica-FLRW}
\end{equation}
where $a(t)$ is the scale factor and $k$ characterizes the spatial curvature: \(k=0\) corresponds to a flat universe, \(k > 0\) to a closed one, and \(k < 0\) to an open geometry.
Adopting this metric (Eq.~\ref{metrica-FLRW}) and treating the cosmic contents as a collection of perfect fluids, one arrives at the Friedmann equation:
\begin{equation}
    H^{2}=\frac{8\pi G \rho}{3} - \frac{k}{a^{2}}\,,
    \label{first_friedmann_equation}
\end{equation}
\noindent
where $H\equiv\dot{a}/{a}$ defines the Hubble parameter, and $\rho$ and $P$ denote the total energy density and pressure of all components, respectively. In this framework, the expansion history of the Universe is governed by its constituents, where each one satisfies the continuity equation:
\begin{equation}
    \dot{\rho} + 3H(\rho + P) = 0\, ,
    \label{continuity_equation}
\end{equation}
whose solution, for barotropic fluids with an equation of state (EoS) $P = \omega \rho$, is given by
\begin{equation} 
    \rho(a) = \rho_0 \exp\left[-3 \int_{a_0}^{a} \frac{1 + \omega(a')}{a'} da' \right],
    \label{general_rho_solution}
\end{equation}
where $\rho_0$ is the present-day density, and, in general, quantities with subscript ‘0’ are evaluated at the current epoch ($a=1$). For constant $\omega$, this expression simplifies to the familiar power-law dependence, whereas for a time-varying $\omega(a)$, one can introduce a function $f$ such that $\rho_{de}(a)=\rho_{de, 0}f(a)$. The function $f(a)$ encodes the behavior of general EoS parameterizations, enabling the dark energy density to evolve in dynamical scenarios.

For a universe containing radiation ($r$), matter ($m$), spatial curvature ($k$), and a dark energy component ($de$), the Friedmann equation (Eq.~\ref{first_friedmann_equation}) can be rewritten as
\begin{equation}\label{Hubble_Omega}
   \frac{H^{2}}{H_{0}^{2}} = \Omega_{r,0}(1+z)^{4} + \Omega_{m,0}(1+z)^{3} + \Omega_{k,0}(1+z)^{2} + \Omega_{de,0} f(z),
\end{equation}
where we have introduced the critical density $\rho_{\text{cr}} = \frac{3 H^{2}}{8 \pi G}$ to define the dimensionless density parameters for each component as $\Omega_{i} = \rho_{i} / \rho_{\text{cr}}$.
Because the radiation contribution becomes negligible at late times, we will ignore it from this point onward.

In dynamical DE scenarios, the EoS parameter $\omega_{\rm de}$ deviates from $-1$ and  evolves with time. Among various possible parameterizations, a frequently used strategy is to expand $\omega_{\rm de}(a)$ in a Taylor series around $a = 1$:
\begin{equation}
    \omega_{de}(a) = \sum^{N}_{j=0} (1-a)^{j} \omega_{j}\,,
    \label{Taylor_cpl}
\end{equation}
where each coefficient $\omega_j$ represents the contribution of the $j$-th order term in the expansion. In this analysis, we keep only the lowest-order terms, which already offer a convenient framework to investigate departures from $\Lambda$CDM; this same approach can, however, be straightforwardly extended to more elaborate models.

\subsection{$\Lambda$CDM model}

The $\Lambda$CDM scenario constitutes the most straightforward dark energy model, as it does not introduce any extra degrees of freedom into Eq.~\ref{Hubble_Omega}. In this framework, dark energy is described by a cosmological constant with
$\omega_{de}=-1$, implying $f(z)=1$. 
The parameters to be optimized are the matter density $\Omega_{m,0}$, the curvature $\Omega_{k,0}$, and the Hubble constant $H_{0}$, while the dark energy density is determined by the relation $\Omega_{de,0}=1-\Omega_{m,0}-\Omega_{k,0}$.

\subsection{$\omega_0$CDM }

This model represents the zeroth-order truncation of Eq.~\ref{Taylor_cpl}, obtained by taking a constant equation-of-state parameter $\omega_{de}=\omega_0$. For the universe to undergo accelerated expansion, $\omega_0$ must fulfill $\omega_0<-1/3$, with the specific case of $\omega_0=-1$ that reproduces the cosmological constant. Values $\omega_0 \geq -1$ characterize the so-called quintessence-like regime, whereas $\omega_0 < -1$ corresponds to the phantom regime \cite{copeland2006dynamics}.
Within the $\omega_0$CDM framework, the solution of Eq.~\ref{general_rho_solution} expressed as a function of redshift is
\begin{equation}
    f(z)=(1+z)^{3(1+\omega_{0})} \,.
\end{equation}
As a result, Eq.~\ref{Hubble_Omega} contains one extra free parameter ($\omega_{0}$) in comparison with the $\Lambda$CDM model.

\subsection{CPL}
The Chevallier-Polarski-Linder (CPL) parametrization~\cite{scherrer2015mapping} corresponds to the first-order truncation of the Taylor expansion in Eq.~\ref{Taylor_cpl} and introduces a linearly varying EoS:
\begin{equation}
    \omega_{\text{CPL}}(a)=\omega_{0}+\omega_{1}(1-a),\quad 
    \omega_{\text{CPL}}(z)=\omega_{0}+\omega_{1}\frac{z}{1+z}\,.
    \label{cpl_scale_factor}
\end{equation}
\noindent
In the early Universe ($a\rightarrow 0$), the EoS tends to $\omega_{de}=\omega_{0} + \omega_{1}$, whereas at the present epoch ($a=1$) it becomes $\omega_{de}=\omega_0$. The parameter $\omega_1$ quantifies the slope of $\omega_{\text{CPL}}$ at $z=0$, and its sign determines whether the EoS increases or decreases with redshift~\cite{Pantazis:2016nky}. From Eq.~\ref{general_rho_solution}, the corresponding dark energy density evolves as
\begin{equation}
    f(z)=(1+z)^{3(1+\omega_{0}+\omega_{1})}
    \exp\left(\frac{-3\omega_{1}z}{1+z}\right)\, .
\end{equation}
Thus, the CPL scenario is specified by two additional free parameters, $\omega_{0}$ and $\omega_{1}$ (the latter is often denoted by $\omega_a$).

\section{Parameter estimation} \label{Sec:paraeter_estimation}

To identify the cosmological model that most accurately reproduces the observational data, we begin by determining the parameter set $\theta_i$ that maximizes the likelihood function $\mathcal{L}(\theta_i)$. Assuming that the observational errors follow a Gaussian distribution, the likelihood can be written as \cite{padilla2021cosmological, heavens2009statistical, verde2010statistical}:
\begin{equation*}
\mathcal{L}({\theta_i}) \propto \exp\left(-\frac{1}{2} \chi^2({\theta_i})\right),     
\end{equation*}
where 
\begin{equation}
\chi^2({\theta_i}) = ({D}^{\text{th}} - {D}^{\text{obs}})^T {C}^{-1} ({D}^{\text{th}} - {D}^{\text{obs}}),
\label{likelihood_equation}    
\end{equation}
measures the mismatch between theoretical predictions ${D}^{\text{th}}$, which depend on the cosmological parameters $\theta_i$, and  the observational data vector ${D}^{\text{obs}}$; $C$ is the covariance matrix associated with the data.

This parameter inference task can be cast as an optimization problem \cite{prasad2012cosmological}. In this work, we employ the Particle Swarm Optimization (PSO) algorithm to carry out this optimization. While PSO is well suited to locating the minimum of $\chi^2$, it does not, by itself, yield estimates of parameter uncertainties or confidence regions around $\theta_{i0}$. To characterize these, we approximate the log-likelihood as Gaussian in the vicinity of its maximum:
\begin{equation}
   \ln \mathcal{L}(\theta_i)= \ln \mathcal{L}(\theta_{i0})  - \frac{1}{2} \Delta \theta_{i} H_{ij} \Delta \theta_{j},
    \label{likelihood_expression}
\end{equation}

\noindent where $\Delta \theta_i=\theta_i -\theta_{i0}$ and $H_{ij}$ is the Hessian matrix of second derivatives of the negative log-likelihood, evaluated at the best-fit point. Within this Gaussian approximation, the inverse of the Hessian serves as an estimator for the parameter covariance matrix, from which one can derive confidence intervals and parameter correlations. Consequently, this approach provides a local Gaussian approximation to the likelihood, yielding symmetric confidence intervals. Strongly non-Gaussian posterior distributions are more accurately characterized through full posterior sampling or more complex approximation methods, i.e. \cite{Amendola:2020qkb}.\\
  
In addition to obtaining best-fit parameters and their uncertainties, we also compare the relative performance of different models by examining both their goodness of fit and their complexity. For this purpose, we adopt the Akaike~\cite{Akaike:1974vps} and Bayesian~\cite{Schwarz:1978tpv} Information Criteria (AIC and BIC), which score models by combining the maximum likelihood with a penalty for complexity. The AIC penalizes the number of free parameters $k$, while the BIC further incorporates the total number of data points $N$, namely:
\begin{equation}
    AIC=-2\ln \mathcal{L}_{\text{max}} + 2k,
    \label{AIC_criteria}
\end{equation}
\begin{equation}
    BIC=-2\ln \mathcal{L}_{\text{max}}+ k \ln N.
    \label{BIC_criteria}
\end{equation}
Lower AIC and BIC values correspond to models that are preferred according to their respective selection criteria.

To carry out our analysis, we employed the set of theoretical models described in the previous section together with different combinations of observational datasets. These datasets offer complementary constraints and thereby enable a consistent and unbiased comparison between models and observations.
\noindent The observational samples used in this study are:
\begin{itemize}[leftmargin=*, itemsep=1.5ex]
  \item \textbf{Type Ia Supernovae.}
     Owing to their well-standardized luminosities, SNe Ia serve as highly effective distance indicators. In this work, we use the most recent compilation sample of Union3:
    \begin{itemize}[leftmargin=*, itemsep=1.5ex]

        \item Union3 \cite{Rubin:2023ovl} is a compilation of 2,087 SNe Ia drawn from 24 different datasets, providing a large and heterogeneous sample.
        In contrast to earlier Union analyzes, it adopts a Bayesian inference framework instead of a frequentist approach, utilizes revised calibration procedures, and implements a new selection strategy. 
    \end{itemize}
    \item \textbf{Baryonic acoustic oscillations.}
   BAO act as cosmological standard rulers, arising from sound waves in the early universe that left a characteristic scale imprinted on the large-scale matter distribution. Measuring this scale with different tracers of the matter field makes BAO a powerful probe of the cosmic expansion history.  
    \begin{itemize}[leftmargin=*, itemsep=1.5ex]
        \item DESIBAO \cite{DESI:2024mwx}, based on the DR1 data release of the DESI survey, delivers precise BAO measurements over the redshift interval \(0.1<z<4.16\), using more than 6 million tracers, including galaxies, quasars, and the Lyman-\(\alpha\) forest. 
    \end{itemize}
\end{itemize}

We therefore constrain the cosmological parameters of several dark energy scenarios using DESI data alone and the combined dataset DESI + Union3.

\section{Particle Swarm Optimization} 
\label{Section_PSO}

Particle Swarm Optimization is a stochastic, population-based method originally inspired by the social interactions of birds in a flock, aiming to mimic the (seemingly) random collective motion as they search for food. In this approach, each particle in the swarm corresponds to a candidate solution, modeled as a computational agent that moves through a multidimensional search space \cite{engelbrecht2007computational}. A central feature of the algorithm is that both individual behavior and group dynamics jointly determine overall performance. In other words, the success of one particle can affect its neighbors and, in turn, guide the entire swarm toward an optimal region: \textit{all particles benefit from the synergy among members}. 
Effective performance relies on communication, shared information, accumulated experience, and dynamic adaptation.

Since its introduction in 1995, Particle Swarm Optimization has been extensively investigated and progressively refined. In this work, we outline the fundamental concepts and provide a concise, general description to convey the core idea of the algorithm, and then justify its main hyperparameters and, consequently, its implementation; for a comprehensive discussion, see \cite{engelbrecht2007computational, kennedy1995particle, skokos2005particle, clerc2010particle, shi2001particle}.

\subsection{Particle Dynamics} 
\label{particle_dynamics}

Let $\vec{x}_{i}(t)$ denote the position vector of the $i$-th particle in the swarm, located in an $n$-dimensional search space at time $t$. To enable communication and thus information sharing, the particle’s position is updated by assigning a velocity $\vec{v}_{i}$ to its current state:
\begin{equation}
    \vec{x}_{i}[t+1]=\vec{x}_{i}[t] + \vec{v}_{i}[t+1].
    \label{position_eq_vec}
\end{equation}
The objective, or \emph{fitness function $f$}, given by $f:U \rightarrow V$ with $U \subseteq \mathbb{R}^{n}$ and $V \subseteq \mathbb{R}$, assesses the quality of each position and consequently governs the subsequent moves.

\noindent
The velocity vector is the central element that guides the optimization and is typically decomposed into three components:
\begin{equation} \label{eq:velocity}
    \vec{v}_{i} = \vec{v}_{\rm intertia} + \vec{v}_{\rm cognitive} + \vec{v}_{\rm social}.
\end{equation}

\begin{itemize}
    \item The \textbf{Inertia component, $\vec{v}_{i}[t]$}, regulates the momentum of particle $i$ by scaling the influence of its previous velocity on the new one, thereby avoiding abrupt changes of direction: \textit{the particle retains memory}.  
    \item The \textbf{Cognitive component, $(\vec{p}_{i}[t]-\vec{x}_{i}[t])$}, measures how well particle $i$ has performed based on its own search history. 
    This term is often referred to as the "nostalgia" component: \textit{the inclination to return to the most favorable position encountered so far}, 
    since it is proportional to the distance between the current position of particle $i$ and its \emph{personal best} position $\vec{p}_{i}=(p_{i1}, p_{i2},...,p_{in})$, or pbest$^i$, defined as
    \begin{align}&\underset{k=0,...,t-1,t}{\operatorname{min}}\;f(\vec{x}_{i}[k])=f(\vec{p}_{i}[t]).
    \end{align}
    \item The \textbf{Social component} characterizes the performance of particle $i$ relative to that of its neighbors.
    It reflects \textit{a social inclination of individuals to imitate the success of their nearby peers.}
    The neighborhood topology determines the social structure of the algorithm.
    Common choices in standard PSO include the star topology, which yields the \emph{Global Best} PSO, or \textit{Gbest}, and the ring topology, which leads to the \emph{Local Best} PSO, or \textit{Lbest}.
\end{itemize}

\subsection{Global Best PSO}

In the \emph{Gbest} PSO, the neighborhood of each particle is defined as the complete swarm, and this is implemented using a Star topology; the left panel of Figure~\ref{star_topology} depicts this kind of topology. 
In this configuration, every particle is linked to all others; thus, at each iteration, the whole swarm is drawn toward the best solution found so far.

\begin{figure}[t]
\centering
    \includegraphics[width=0.35\linewidth]{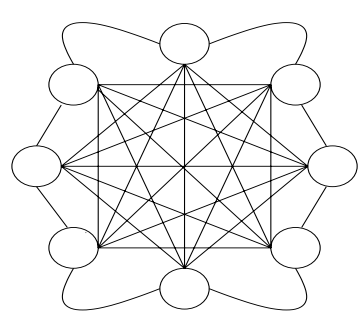}
    \includegraphics[width=0.35\linewidth]{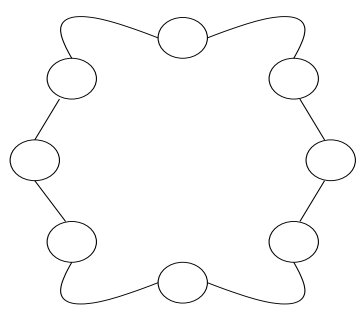}
    \caption{Left panel: star topology, where the size of the swarm corresponds to $N=8$ and all particles are connected. 
    Right panel: ring topology, with $N=8$ and $n_{\mathcal{N}}=2$ neighbors. 
    Obtained from \cite{engelbrecht2007computational}.}
    \label{star_topology}
\end{figure}

For the \textit{Gbest} PSO, the social term is expressed as
\begin{equation}
\left(\vec{p}_{g}[t]  - \vec{x}_{i}[t]  \right),    
\end{equation}
where $\vec{p}_{g}$ denotes the \emph{global best}, or \emph{gbest}, position among all $N$ particles in the swarm. 
In a minimization context, this is the position associated with the smallest fitness value, defined as
\begin{align}&\underset{j=1,...,N}{\operatorname{min}}\;f(\vec{p}_{j}[t])=f(\vec{p}_{g}[t]).
\end{align}

Consequently, the three components of the velocity update (Eq.~\ref{eq:velocity}) for the \textit{Gbest} PSO reduce to
\begin{equation}
    \vec{v}_{i}[t+1]= w \vec{v}_{i}[t] + r_{1}c_{1}\left(\vec{p}_{i}[t]-\vec{x}_{i}[t]\right) + 
    r_{2}c_{2}\left(\vec{p}_{g}[t]  - \vec{x}_{i}[t]  \right).
    \label{velocity_eq_vec}
\end{equation}

The pseudo-code, shown in Algorithm~\ref{algo1}, outlines the procedure to implement \textit{Gbest} PSO (adapted from \cite{mratinkovic2019illustrated}).

\begin{algorithm}[H]
\caption{\textit{Global Best algorithm.}}
\label{algo1}
  \begin{algorithmic}[1]
 %
    \Require{\textit{initialize}: Size of the swarm $N$. Maximum iterations number $max\_iter$.}
\While{ not (end condition)}
\For{$i=\ (1 \ to \ n)\ $}
    \State $\vec{x}_{i}, \vec{v}_{i}  \gets$ \ initialize $i$th particle position and velocity.
    \If{$f(\vec{x}_{i}) < f(\vec{p}_{i}) $}
          \State $\vec{p}_{i}=\vec{x}_{i}$ 
    \EndIf \State{\textbf{end if}}
    \If{ $f(\vec{p}_{i}) < f(\vec{g}))$} 
        \State $\vec{p}_{i}=\vec{g} \gets $ \ set best global position.
    \EndIf \State{\textbf{end if}}
\EndFor \State{\textbf{end for}}
\For{$i=\ (1 \ to \ n)\ $}
    \State $\vec{v}_{i}  \gets$ \ update $i$th particle velocity using
    \State $\vec{x}_{i}(t+1)=\vec{x}_{i}(t)+\vec{v}_{i}(t+1)$
\EndFor
    \State{\textbf{end for}}
\EndWhile
\State{\textbf{until} the condition is satisfied}
   \end{algorithmic}
\end{algorithm}

For both variants, \textit{Gbest} PSO (Eq.~\ref{velocity_eq_vec}) and \textit{Lbest} PSO (Eq.~\ref{velocity_eq_vec_Lbest}), the inertia weight $w$ mediates the trade-off between global exploration and local exploitation.
The positive coefficients $c_{1}$ and $c_{2}$, termed the \emph{cognitive} and \emph{social} parameters, respectively, modulate the search behavior of the swarm. The cognitive parameter $c_{1}$ enhances the particle’s individual exploration, whereas the social parameter $c_{2}$ governs the influence of the collective swarm search.
To incorporate randomness into the algorithm, $r_{1}$ and $r_{2} \in \mathcal{U}(0,1)$ are random variables uniformly distributed in the interval $[0,1]$.

\subsection{Local Best PSO}

In the global best configuration, particles generally converge rapidly; however, this occurs at the expense of diversity and makes the swarm more prone to getting trapped in local optima \cite{kennedy2001morgan}. To enhance the exploration of the search space and mitigate stagnation, the local version of PSO, known as \textit{Lbest}, was proposed. The \textit{Lbest} PSO variant employs small neighborhood structures in its implementation to encourage more effective search behavior, albeit with a reduced convergence speed \cite{mohanty2012particle}. 

Consider a swarm of particles $S=\{x_{1},x_{2},...,x_{N} \}$, where $N$ denotes the total number of particles. The neighborhood of the $i$th particle $x_{i}$ is defined as:
\begin{equation}
    B_{i}=\{x_{n_{1}}, x_{n_{2}},...,x_{n_{s}}\},
\end{equation}

\noindent 
where $\mathcal{N}=\{n_{1}, n_{2},...,n_{s} \}\subseteq \mathcal{I}=\{1,2,...,N \}$ is the set of indices corresponding to the neighbors of $x_{i}$. A frequently adopted structure for the \textit{Lbest} PSO is the Ring topology, in which each particle is connected to its $n_{\mathcal{N}}$ closest neighbors. The particular case with $n_{\mathcal{N}}=2$ is depicted in the right panel of Figure~\ref{star_topology}.
Note that the \textit{Gbest} algorithm can be regarded as a special instance of the \textit{Lbest} PSO in which $n_{\mathcal{N}}=n_{s}$.
\\

Therefore, the structure of the velocity for the \textit{Lbest} PSO becomes 
\begin{equation}
    \vec{v}_{i}[t+1]= w \vec{v}_{i}[t] + r_{1}c_{1}\left(\vec{p}_{i}[t]-\vec{x}_{i}[t]\right) + 
    r_{2}c_{2}\left(\vec{p}_{gi}[t]  - \vec{x}_{i}[t]  \right),
    \label{velocity_eq_vec_Lbest}
\end{equation}

\noindent where, $\vec{p}_{gi}$ represents the best position found by the neighborhood of the $i$-th particle.
The pseudo-code, Algorithm~\ref{algo_2},  shows the steps to implement \textit{Lbest} PSO (from~\cite{mratinkovic2019illustrated}).

\begin{algorithm}[H]
   \caption{\textit{Local Best algorithm.}}
   \label{algo_2}
  \begin{algorithmic}[1]
    \Require{\textit{initialize}: Size of the swarm $N$. Number of neighborhoods $k$. Maximum iterations number $max\_iter$}.
\While{ not (end condition)}
\For{$i=\ (1 \ to \ n)\ $}
      \State $\vec{x}_{i}, \vec{v}_{i}  \gets$ \ initialize $i$th particle position and velocity.
\If{$f(\vec{x}_{i}) < f(\vec{p}_{i}) $}
          \State $\vec{p}_{i}=\vec{x}_{i}$ 
          \EndIf
          \State{\textbf{end if}}
          \If{ $f(\vec{p}_{i}) < f(\vec{g}_{i}))$} 
          \State $\vec{p}_{i}=\vec{g}_{i} \gets $ \ set best neighborhood position.
          \EndIf
          \State{\textbf{end if}}
    \EndFor
\State{\textbf{end for}}  
\For{$i=\ (1 \ to \ n)\ $}
\State $\vec{v}_{i}  \gets$ \  update $i$th particle velocity using Eq.~\ref{velocity_eq_vec_Lbest}.
\State $\vec{x}_{i}(t+1)=\vec{x}_{i}(t)+\vec{v}_{i}(t+1)$
\EndFor 
\State{\textbf{end for}}
\EndWhile
\State{\textbf{until} the end condition is satisfied}
\end{algorithmic}
\end{algorithm}

In addition to the ring topology, a few examples of local structures are the following \cite{kennedy2002population, kennedy2006neighborhood, miranda2020pyswarms}:  
\begin{itemize}
    \item The \textbf{Pyramid} structure forms a 3-dimensional wireframe triangle.
    \item In the \textbf{Von Neumann} neighborhood, the population is arranged in a rectangular grid structure. 
    \item In the \textbf{Random} structure, as its name states, particles are connected to their $n_{s}$ random particles.
\end{itemize}

\subsection{General Aspects of PSO}

\subsubsection{Initial Conditions}
%
A typical strategy for selecting the particles’ initial positions is to sample them uniformly at random within the hypercube domain defined by $\vec{x}_{\rm min}$ and $\vec{x}_{\rm max}$, that is, $\vec{x}_i(0)\in \mathcal{U}(\vec{x}_{\rm min},\vec{x}_{\rm max})$, or more explicitly:
\begin{eqnarray}
    \vec{x}_i(0)= \vec{x}_{{\rm min},i} + r_i(\vec{x}_{{\rm max},i}- \vec{x}_{{\rm min},i}),
\end{eqnarray}
with $ r_i\sim \mathcal{U}(0,1)$. In addition, the personal best position can be initialized directly from the initial position:
\begin{equation}
    \vec{p}_{i}(0)= \vec{x}_{i}(0).    
\end{equation}
If the particles are assumed to be initially at rest, their velocities can be set to zero:
\begin{equation}
    \vec{v}_{i}= \vec{0}.
\end{equation}

The velocities may also be initialized randomly in a way analogous to the positions; however, this must be done carefully. Various approaches for initializing both the position and the velocity of the particles have been proposed and analyzed in the literature \cite{engelbrecht2012particle, helwig2008theoretical}.

\subsubsection{Stopping Criteria}

A straightforward stopping rule is to terminate the algorithm after a predefined number of iterations. However, this basic approach can be ambiguous: selecting an arbitrary upper limit for the iterations may lead either to premature convergence (ending the search before a sufficiently good solution is obtained) or, conversely, to unnecessary evaluations of the fitness function, thus increasing the computational cost.

A more suitable way to determine convergence is to stop when an acceptable solution has been reached, namely when
\begin{equation}
    || f(\vec{x}_{i}[t]) - f(\vec{x}^{*}[t])|| < \varepsilon,
    \label{error_condition}
\end{equation}

\noindent 
where $f(\vec{x}^{*}[t])$ denotes the fitness function evaluated at the position $\vec{x}^{*}[t]$ of the best global minimum found so far, and $\varepsilon$ is a predefined positive tolerance level \cite{parsopoulos2010particle}. If no improvement is observed over several consecutive iterations and condition (\ref{error_condition}) remains satisfied throughout this interval, the PSO can be terminated. An appropriate choice of both the accuracy threshold and the required number of consecutive iterations helps ensure that the optimal value is correctly identified.

\subsubsection{Acceleration Coefficients} 
\label{acceleration_coeff_sec}

The choice of acceleration parameters depends on the specific goals of the search. Early formulations of PSO typically used identical values for the cognitive and social components, i.e., $c_{1}=c_{2}$. When strong global exploration is desired, both $c_{1}$ and $c_{2}$ should be set to relatively large values. In contrast, smaller values are preferable when focusing on local refinement near optimal solutions. Furthermore, by considering the geometry of the objective function, it is possible to impose additional preferences, such as $c_{1}<c_{2}$ for convex unimodal objective functions~\cite{parsopoulos2010particle}. To guaranty convergence, the acceleration coefficients must satisfy
\begin{equation}
    c_{1}+c_{2} \leq 4.
\end{equation}
If this condition is violated, particle velocities and positions can become unbounded and diverge \cite{engelbrecht2007computational}.

\subsubsection{Velocity Clampling and Inertia Weight} 
\label{inertia_weight_sec}

Although the acceleration coefficients tune the search capability via the \textit{pbest} and \textit{gbest} values, the velocity itself can still become unbounded, allowing particles to make excessively long moves that may cause the swarm to diverge. To address this, the earliest PSO variants introduced boundary constraints so that the velocity is restricted to remain within the search space. By defining a maximum allowable velocity \(v_{\text{max}}>0\), the particle position updates are controlled through the velocity magnitude:
\[
  v_{i}[t+1] = 
  \left \{
    \begin{aligned}
      v_{i}'[t+1]  &,\ \quad \text{if} \ v_{i}'[t+1] < v_{\rm max},\\
      v_{\rm max} &,\ \quad \text{if} \  v_{i}'[t+1] > v_{\rm max},
    \end{aligned}
  \right .
\]
where \(v_{i}'\) is computed either from Eq.~(\ref{velocity_eq_vec}) or from Eq.~(\ref{velocity_eq_vec_Lbest}).

While the maximum velocity bound regulates the global exploration behavior of the particles, it also introduces additional challenges. If \(v_{\text{max}}\) is chosen too small, more iterations are required, and the swarm is more likely to become trapped in local optima. Conversely, if \(v_{\text{max}}\) is too large, particles may skip over promising regions and move directly into unproductive areas. Several guidelines have been proposed for choosing \(v_{\text{max}}\); for instance, \cite{engelbrecht2007computational} suggests
\begin{equation}
    v_{\text{max}}=\delta({x_{\text{max}}-x_{\text{min}}}),
\end{equation}

\noindent where \([x_{\text{min}}, x_{\text{max}}]\) denotes the domain bounds, and \(\delta \in (0,1]\). The choice of \(\delta\) is problem-dependent; for example, setting \(\delta=0.5\) implies that the largest possible move of a particle corresponds to half of the search space.

The \emph{inertia weight} parameter, \(w\), can mitigate the need for a finely tuned maximum velocity \(v_{\text{max}}\). This parameter is widely regarded as having the strongest impact on PSO performance, as it governs the balance between exploration and exploitation and typically requires less delicate tuning~\cite{wang2018particle}. In particular, small values of \(w\) promote local exploitation, so that the cognitive and social terms dominate the position updates, whereas large values of \(w\) enhance exploration by increasing swarm diversity. However, under velocity clamping, if \(w \geq 1\), the velocities grow over time and the swarm diverges, while for \(w<1\) the particles gradually slow down and their velocities eventually vanish.

As with velocity clamping, an appropriate setting of the inertia weight is problem-specific. Nevertheless, several strategies have been proposed to obtain a desirable behavior of \(w\), and its selection must be coordinated with that of \(c_1\) and \(c_2\).
Some examples include the following

\begin{itemize}
    \item \textbf{Static inertia weight}. As the term indicates, the constant $w$ is kept unchanged throughout all swarm iterations. It has been demonstrated that if the following condition is satisfied:
    \begin{equation}
    w > \frac{1}{2} (c_{1} + c_{2}) -1,
    \label{inertia_weight_condition}
    \end{equation}
    then convergence of the algorithm is ensured. Shi and Eberhart \cite{shi1998parameter} recommended choosing $w$ within the range [0.9, 1.2]; otherwise, the swarm may fail to converge and instead diverge or enter into oscillatory (cyclic) patterns \cite{engelbrecht2007computational}. 
    
    \item \textbf{Linear time decreasing}. In this strategy, $w$ varies over time. It begins with a relatively high value and is gradually reduced to a lower value over the course of the iterations $t$. This linear reduction enables broader global exploration at early stages, followed by more intensive local refinement near the end \cite{shi1999empirical}. The linear schedule is defined as:
    \begin{equation}
    w[t]=\left( w[0] - w[n_{t}] \right)\frac{n_{t}-t}{n_{t}} + w[n_{t}],
    \end{equation}
    where $w[0]$ and $w[n_{t}]$ are the initial and final inertia weights, respectively, with $w[0] > w[n_{t}]$, and $n_{t}$ denoting the maximum number of iterations. Typical choices for $w[0]$ and $w[n_{t}]$ are $0.9$ and $0.4$, respectively.

    \item \textbf{Exponential decreasing}. The scheme proposed in \cite{lu2015generalized} is another time-varying approach for $w$ that aims to trade off global exploitation and local exploration. Here, the inertia weight decays exponentially as $t$ increases. It is defined by:
    \begin{equation}
    w[t]=w[n_{t}] + (w[0] - w[n_t])\cdot\exp\left(-\frac{ct}{n_{t}} \right),
    \end{equation}
    where $c$ is a positive control parameter that determines the rate at which $w$ converges.
\end{itemize}

Given the \textit{Gbest} and \textit{Lbest} algorithms, a natural question that arises is which one is more suitable for a particular optimization task. The key difference between them lies in the neighborhood topology and, consequently, in how information is shared within the swarm. 
In~\cite{engelbrecht2013particle}, the authors evaluated both approaches on several optimization benchmarks (test functions) and found that \textit{Gbest} tended to converge too quickly to local optima, whereas \textit{Lbest} converged more slowly but exhibited stronger exploration capabilities. They also reported that \textit{Gbest} and \textit{Lbest} achieved similarly accurate solutions for a comparable number of functions. Thus, the choice between \textit{Gbest} and \textit{Lbest} becomes a problem-dependent. Before applying them to cosmological problems, we therefore carried out a set of preliminary tests.

The PSO algorithm has been extensively modified and implemented in many programming languages. In the present study, we integrate the \textbf{PySwarms} Python library, described in \cite{miranda2020pyswarms}, into the \href{https://github.com/ja-vazquez/SimpleMC/blob/master/simplemc/analyzers/PSO_optimizer.py}{SimpleMC} code. PySwarms is advantageous because it is straightforward to implement and offers the flexibility needed to tailor the algorithm to our specific goals. Using this library, we apply the \textit{Gbest} and \textit{Lbest} PSO variants to compare their behaviors. The \textit{Lbest} variant additionally requires specifying two parameters: the number of neighborhoods $k$ and a flag indicating the distance metric, which can be either Euclidean or Absolute.

\begin{figure*}[t]
\captionsetup{font=small}
    \centering
    \makebox[11cm][c]{

\includegraphics[trim = 0mm  0mm 0mm 0mm, clip, width=4.5cm, height=4.5cm]{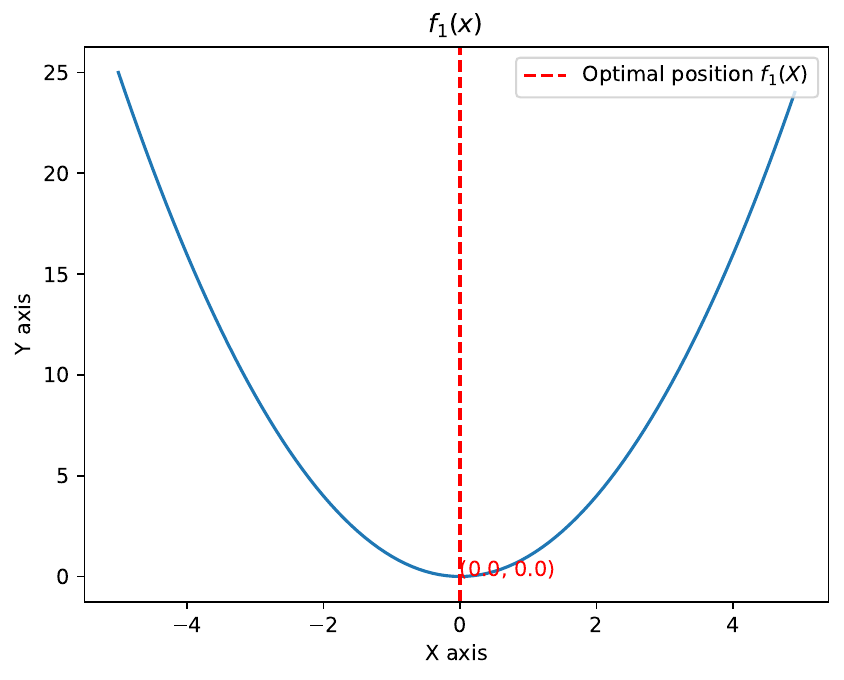}
\includegraphics[trim = 0mm  0mm 0mm 0mm, clip, width=4.5cm, height=4.5cm]{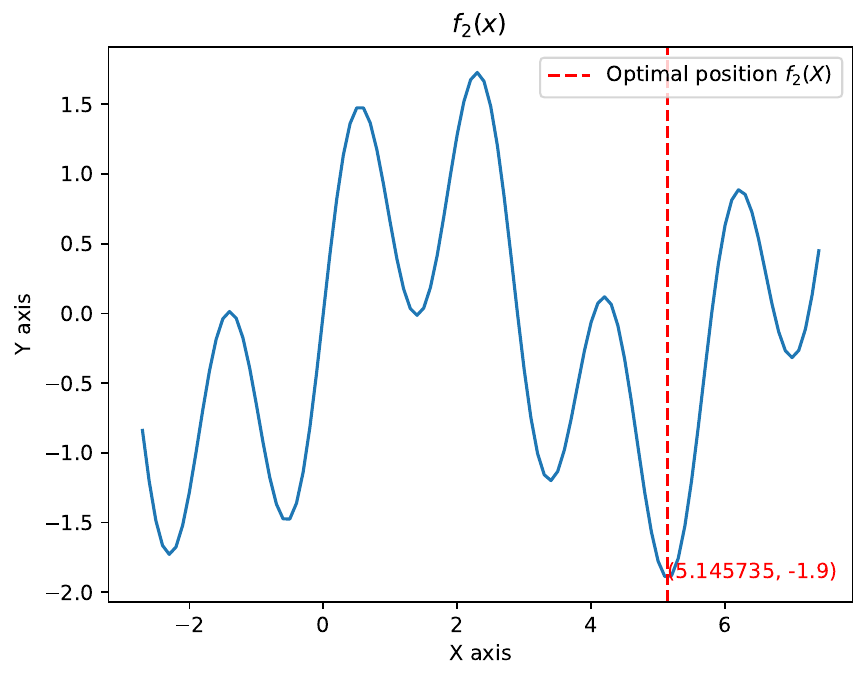}
\includegraphics[trim = 0mm  0mm 0mm 0mm, clip, width=4.5cm, height=4.5cm]{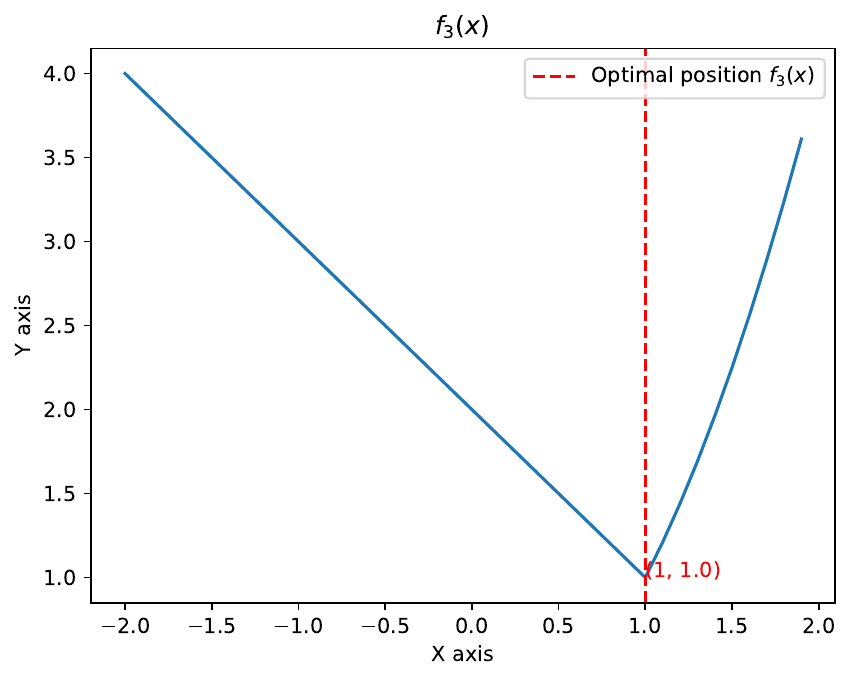}
\includegraphics[trim = 0mm  0mm 0mm 0mm, clip, width=4.5cm, height=4.5cm]{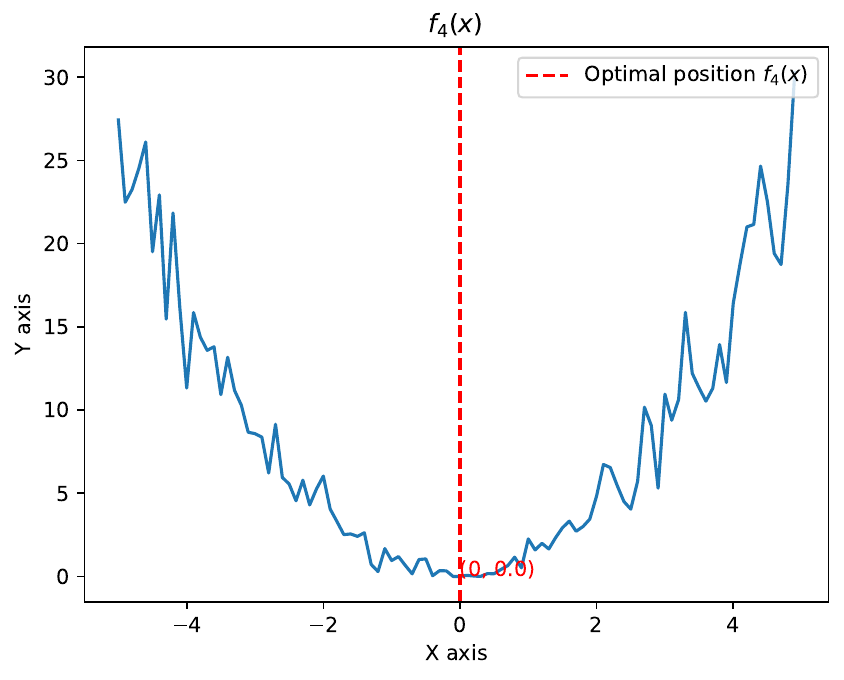}
}
    \makebox[11cm][c]{
\includegraphics[trim = 0mm  0mm 0mm 0mm, clip, width=5cm, height=5cm]{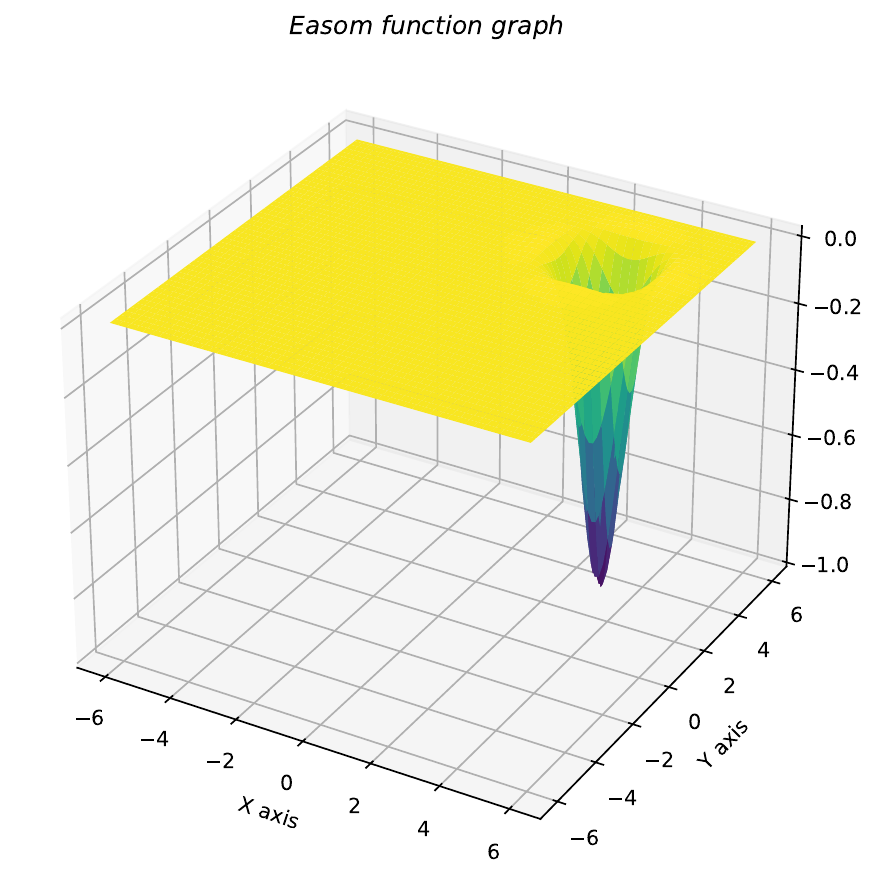}
\includegraphics[trim = 0mm  0mm 0mm 0mm, clip, width=5cm, height=5cm]{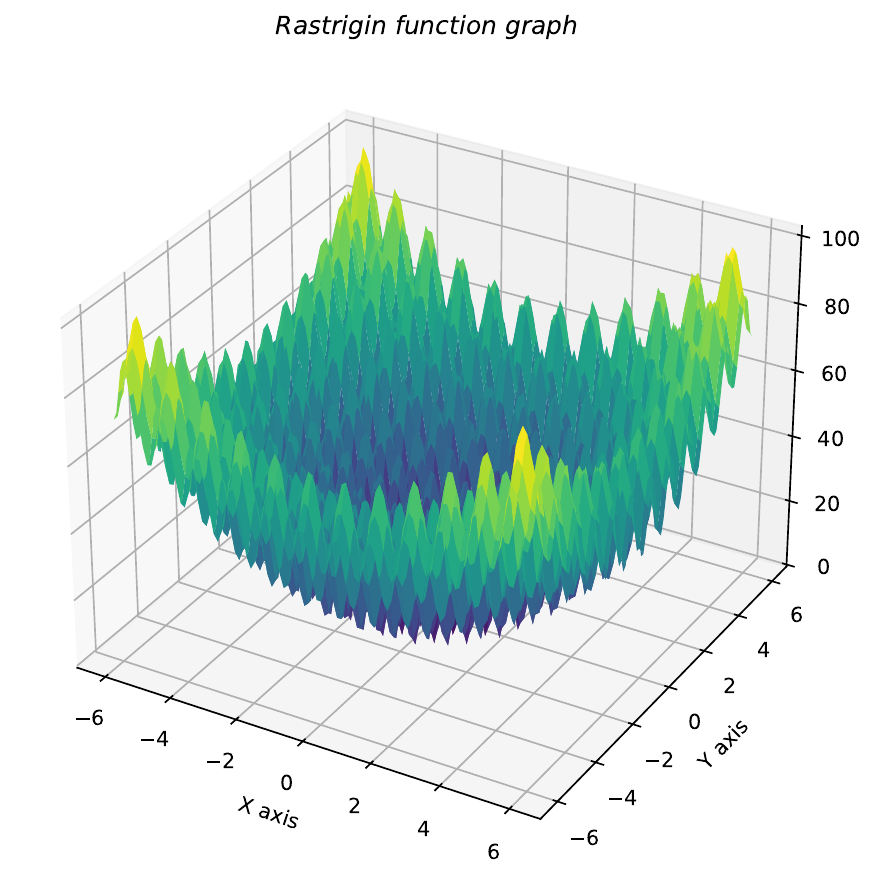}
\includegraphics[trim = 0mm  0mm 0mm 0mm, clip, width=5cm, height=5cm]{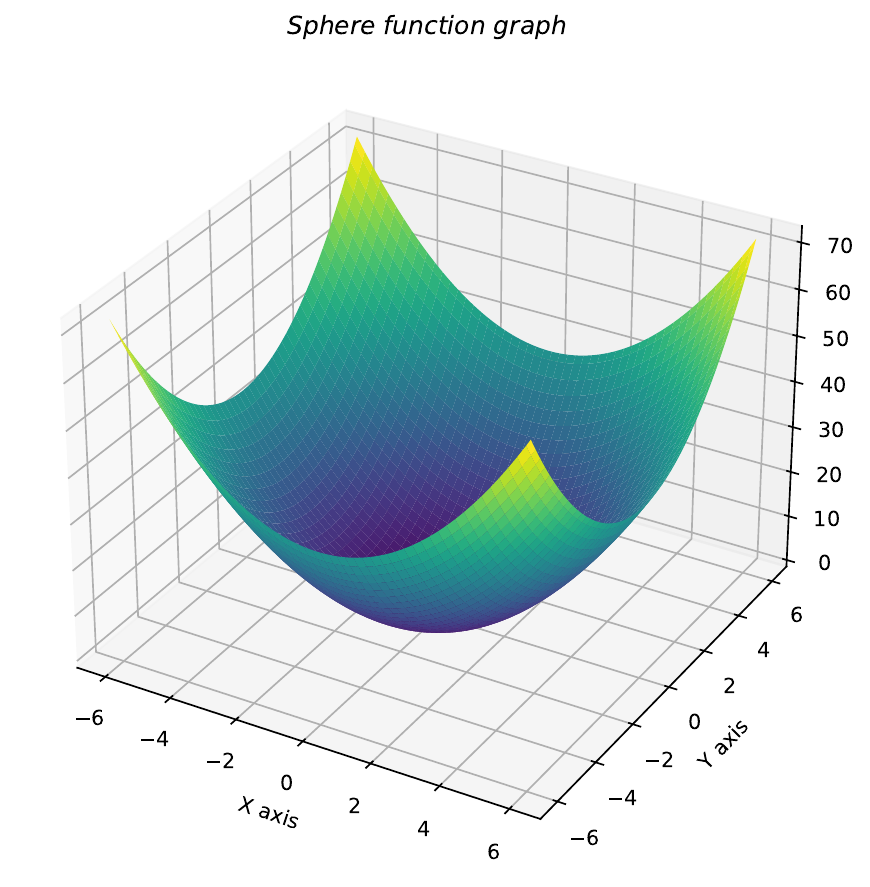}
}
   \caption{Top panel: one dimensional test functions, from left to right: $f_{1}$, $f_{2}$, $f_{3}$ and $f_{4}$.
    Bottom panel: Two dimensional test functions, Easom ($f_{5}$), Rastrigin ($f_{6}$) and Sphere ($f_{7}$).}
   \label{fig:triangleplot_all}
\end{figure*}

\subsection{Testing the Code} \label{testing_code}

The test functions serve as important tools for analyzing an algorithm’s performance and behavior since their true solutions are already known.  
There exists a large variety of test functions with diverse properties, and for some of them, it is considerably harder than for others to locate the global optimum. We concentrate on a small subset: we chose several functions that are multimodal, multidimensional, convex, or that may even exhibit discontinuities. They are shown in Figure~\ref{fig:triangleplot_all} and are defined as follows:
\begin{eqnarray*}
    f_{1}(x)&=&x^{2},\\
    f_{2}(x)&=&\sin(x)+\sin((10/3)x), \\
    f_{3}(x) &=& 
  \left \{
    \begin{aligned}
      x^{2} &,\ \text{si} \ x \geq 1, \\
      2-x &,\ \text{si} \ 
      x <1,
    \end{aligned}
   \right.\\
   f_{4}(x)&=&(x+ (\mathcal{U}(0,1) \cdot 0.3 x))^{2},\\
   f_{5}(\textbf{x}) &=& -\cos(x)\cdot\cos(y) \exp(-(x-\pi)^{2}-(y-\pi)^{2}), \\
    f_{6}(\textbf{x}) &=&20 + (x^{2}-10\cos(2\pi x)+y^{2}-10\cos(2\pi y)),\\
    f_{7}(\textbf{x}) &=&x^{2} + y^{2}.
\end{eqnarray*}

Table~\ref{one_and_two_dim_test_func_results} presents the chosen PSO hyperparameters—selected according to the criteria discussed above—together with the outcomes obtained with the \textit{Gbest} and \textit{Lbest} variants. The results closely reproduce the true position and fitness values, as can be seen by comparing the last three columns.

\begin{table*}[t]
\centering
\resizebox{\textwidth}{!}{
\begin{tabular}{|c|c|ccc|c|c|c|c|c|c|c|}
\hline
\textbf{Function} & \textbf{Struct.} & $c_1$ & $c_2$ & $w$ & \textbf{Iter.} & \textbf{Particles} & \textbf{Neigh. $k$} & \textbf{Distance $p$} & \textbf{Best Position} & \textbf{Best Fitness} & \textbf{Real Pos./Fitness} \\
\hline
\multirow{2}{*}{$f_{1}(x)$} 
  & Gbest & 0.3 & 0.5 & 0.4 & 20 & 10 & \dots & \dots & -6.05$\times 10^{-6}$ & 3.66$\times10^{-11}$ & 0 / 0 \\
  & Lbest & 0.2 & 0.5 & 0.5 & 30 & 20 & 5 & Euclidean & 6.19$\times 10^{-6}$ & 3.84$\times 10^{-11}$ &  \\
  \hline
\multirow{2}{*}{$f_{2}(x)$} 
  & Gbest & 0.7 & 0.3 & 0.9 & 30 & 90 & \dots & \dots & 5.146 & -1.899 & 5.147 / -1.9 \\
  & Lbest & 0.8 & 0.3 & 0.5 & 112 & 20 & 10 & Euclidean & 5.146 & -1.899 &  \\
\hline
\multirow{2}{*}{$f_{3}(x)$} 
  & Gbest & 0.2 & 0.7 & 0.7 & 20 & 25 & \dots & \dots & 0.999 & 1.000 & 1 / 1 \\
  & Lbest & 0.4 & 0.9 & 0.7 & 33 & 10 & 7 & Abs. & 0.999 & 1.000 &  \\
\hline
\multirow{2}{*}{$f_{4}(x)$} 
  & Gbest & 0.5 & 0.5 & 0.9 & 15 & 55 & \dots & \dots & 0.129 & 7.92$\times 10 ^{-7}$ & 0 / 0 \\
  & Lbest & 0.8 & 0.3 & 0.9 & 50 & 10 & 5 & Abs & 0.079 & 2.08$\times 10^{-6}$ &  \\
\hline
\multirow{2}{*}{Easom} 
  & Gbest & 0.3 & 0.9 & 0.9 & 73 & 20 & \dots & \dots & (3.142, 3.142) & -0.999 & ($\pi$, $\pi$) / -1 \\
  & Lbest & 0.5 & 0.8 & 0.7 & 45 & 30 & 9 & Abs. & (3.142, 3.141) & -0.999 &  \\
\hline
\multirow{2}{*}{Rastrigin} 
  & Gbest & 0.8 & 0.4 & 0.9 & 150 & 50 & \dots & \dots & (-4.31$\times10^{-6}$, 3.02$\times10^{-5}$) & 1.85$\times10^{-7}$ & (0,0) / 0 \\
  & Lbest & 0.8 & 0.4 & 0.7 & 200 & 45 & 15 & Euclidean & (-3.27$\times10^{-6}$, 6.31$\times10^{-9}$) & 0.000 &  \\
\hline
\multirow{2}{*}{Sphere } 
  & Gbest & 0.3 & 0.6 & 0.7 & 60 & 30 & \dots & \dots & (5.55$\times$10$^{-7}$, 6.31$\times$10$^{-7}$) & 7.06$\times$10$^{-13}$ & (0,0) / 0 \\
  & Lbest & 0.3 & 0.6 & 0.7 & 30 & 20 & 5 & Euclidean & (0.000, -0.000) & 1.99$\times$10$^{-7}$ &  \\
\hline
\end{tabular}
}
\caption{PSO hyperparameters and optimization results for one and two dimensional test functions.}
\label{one_and_two_dim_test_func_results}
\end{table*}
After validating the code with several analytical test functions, we perform an analogous study, but now using a toy data-problem: fitting a straight line to synthetic data.

In this case, we employ the PSO algorithm to infer the parameters $m$ and $b$ of a linear model ($y = mx + b$) by minimizing the chi-squared function $\chi^{2}$ (Eq.~\ref{likelihood_equation}) for a given mock data set. The synthetic data set is composed of 30 points generated from the underlying values $m=2$ and $b=6$, with Gaussian noise added to each point and standard deviations, represented by black dots and red error bars in Figure~\ref{fig:best_fit_line}.
To determine the best-fit parameters, we adopt the static Global Best PSO scheme and explore several combinations of inertia ($w$) and acceleration coefficients ($c_1$ and $c_2$), seeking the configuration that yields the minimum value of the $\chi^2$ function. In line with the previous procedure, the final PSO settings are reported in Table~\ref{straight_line_pso_parameters}.

\begin{table*}[t]
\begin{tabular}{ccccccc}
\multicolumn{7}{c}{Global Best hyperparameters for $\chi^{2}_{\text{line}}$.}   
       \\ \hline
Function    & Parameter $c_{1}$ & Parameter $c_{2}$ & Inertia weight $\omega$ & Iterations & Particles & Boundaries \\ \hline
$\chi^{2}_{\text{line}}$ & 0.1  & 0.5  & 0.9 & 90           & 200          & $[-10,10]\times [-10,10]$       \\ \hline
\end{tabular}
\caption{PSO parameters to optimize the $\chi^{2}$ function for a straight line.}
\label{straight_line_pso_parameters}
\end{table*}

The optimization yields the following results:
the minimum value of the fitness function is $\chi^{2}_{\text{line}} = 28.59$, with a best-fit slope $m = 2.054$ and intercept $b = 6.025$, which are very close to the true values used to generate the data.
The upper panel of Figure~\ref{fig:best_fit_line} displays the best-fitting (blue) line obtained with the PSO-derived parameters $(m,b)$, while the lower panel shows the evolution of the iterations, illustrating how the swarm particles converge toward the minimum.

\begin{figure}
\captionsetup{font=small}
    \centering

\includegraphics[trim = 0mm  0mm 0mm 0mm, clip, width=8.5cm, height=5.5cm]{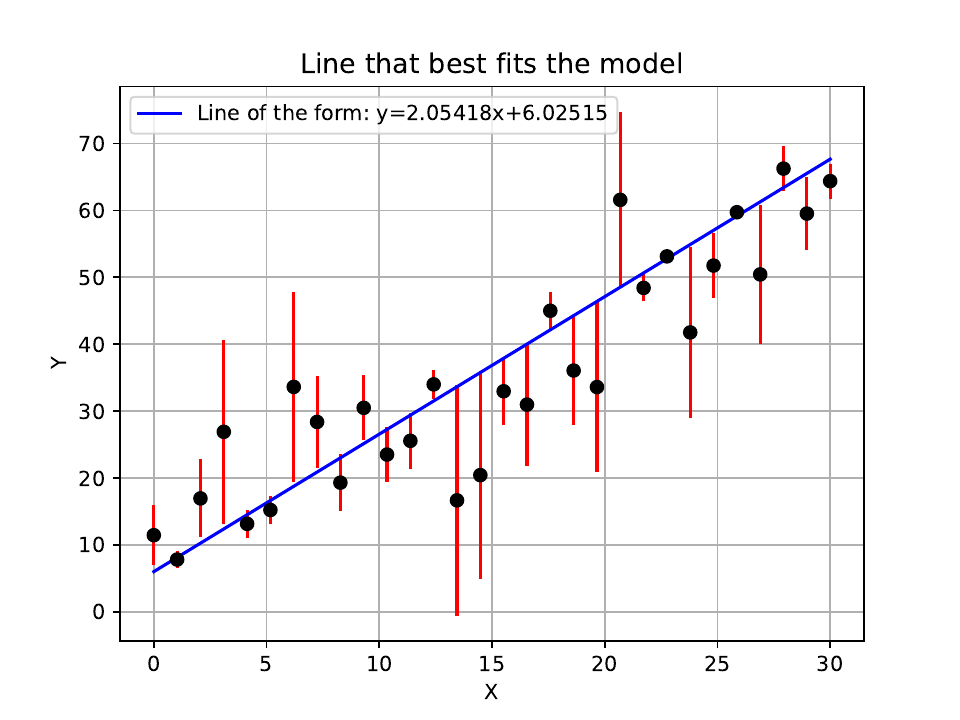}\\
\includegraphics[trim = 0mm  0mm 0mm 0mm, clip, width=8.5cm, height=5.5cm]{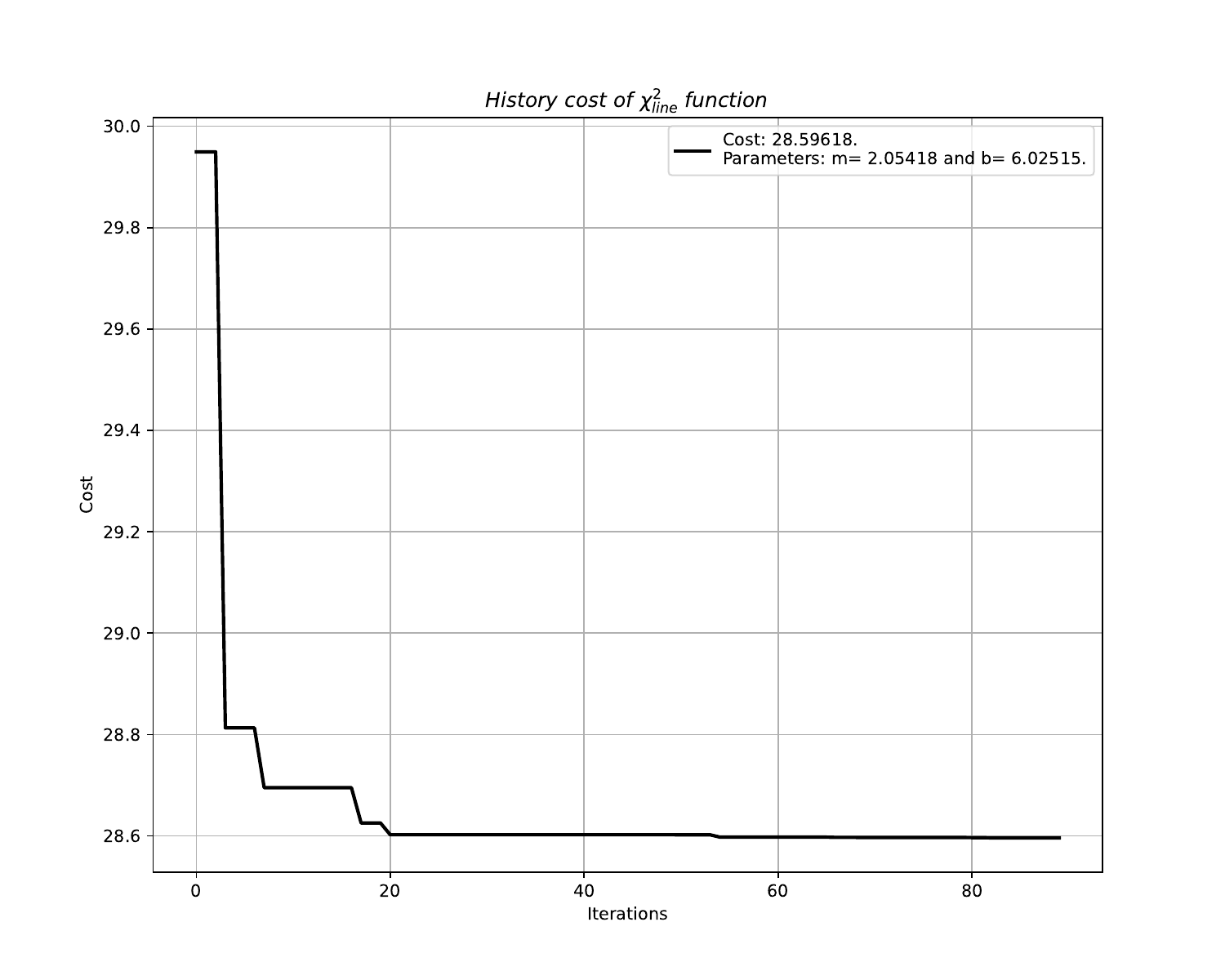}
   \caption{Top: Blue line that best fits the synthetic data using PSO: $m=2.05$ and $b=6.02$.
   Bottom: Cost history of $\chi^{2}_{\text{line}}$ during 90 iterations.}
   \label{fig:best_fit_line}
\end{figure}

\section{Parameter estimation on Dark Energy models}
\label{Seccion-7}

We incorporated the PSO algorithm into a modified version of the \href{https://github.com/ja-vazquez/SimpleMC/blob/master/simplemc/analyzers/PSO_optimizer.py}{SimpleMC} code to minimize the $\chi^2$ function (Eq.~\ref{likelihood_equation}) for several cosmological models: the flat $\Lambda$CDM scenario, $\Lambda$CDM with free spatial curvature (denoted $\Lambda$CDM+$\Omega_k$), and the flat CPL model ($\omega_0\omega_a$CDM). 
The $\Lambda$CDM+$\Omega_k$ model is constrained using only DESI BAO data, whereas the flat $\Lambda$CDM and CPL models are constrained with a combination of DESI and Union3 supernova measurements. 
The set of free cosmological parameters includes $\Omega_m$, $\Omega_k$, the reduced Hubble parameter $h=H_0/100$, and, where applicable, the dark energy equation-of-state parameters $(\omega_0, \omega_a)$. To guaranty a coherent scan of the parameter space, we defined the priors within which the particles can move ($[\vec{x}_{\text{min}}, \vec{x}_{\text{max}}]$) as: $h\in[0.4,0.9]$, $\Omega_m\in[0.1,0.5]$, $\Omega_bh^2\in[0.02,0.025]$, $\Omega_k\in[-0.3,0.3]$,  $\omega_0\in[-2,0]$, and $\omega_a\in[-2,2]$. For comparison, we also integrated into the SimpleMC framework \cite{BOSS:2014hhw}, a modified version of the \textit{dynesty} nested sampling algorithm, a dynamic sampling that efficiently computes the Bayesian evidence while performing parameter estimation \cite{simplemc_github, Speagle:2019ivv}.

As it is well established that the posterior distributions under consideration are unimodal, and to reduce computational cost, the best-fit parameters were determined by using the Global Best PSO scheme, adopting the hyperparameter settings identified as optimal in prior tests: $w=0.9$, $c_1=0.5$, $c_2=0.8$, 100 particles with uniformly random initial positions, and a static configuration, with a maximum of 150 iterations. We verified that different random realizations of the initial swarm consistently converged to the same best-fit solution within the adopted convergence criterion. We also imposed a tolerance criterion of $\varepsilon > 10^{-5}$ \footnote{The selected tolerance was determined according to the minimum difference attained by the MCMC algorithm in the vicinity of its convergence phase.}, stopping the algorithm if the likelihood showed no improvement after, for instance, 10 consecutive iterations. 
Table \ref{tab:parametros_pso} summarizes the resulting best-fit parameters and the corresponding maximum value of the fitness function. The table further lists the AIC and BIC differences with respect to the flat $\Lambda$CDM model; positive values indicate a poorer fit, while negative values favor the extended (beyond-$\Lambda$CDM) model.

\begin{table*}[t]
\begin{tabular}{lcccc cccc}
\toprule
{} & \multicolumn{4}{c}{\textbf{DESI}} &  \multicolumn{4}{c}{\textbf{DESI+Union3}} 
\\
\textbf{Parameter} & \multicolumn{2}{c}{ \textbf{\boldmath{$\Lambda$}CDM}} &  \multicolumn{2}{c}{\textbf{\boldmath{$\Lambda$}CDM+\boldmath{$\Omega_k$}}} & \multicolumn{2}{c}{\textbf{\boldmath{$\Lambda$}CDM}} & \multicolumn{2}{c}{\textbf{CPL }} 
\\
& PSO & MCMC & PSO & MCMC & PSO & MCMC & PSO & MCMC \\ 

\hline
$H_0$           & 68.307  & 68.335   & 66.066 & 66.249 & 68.368 & 68.377 &  67.331 & 67.4651   \\
$\Omega_m$      & 0.2936  & 0.2935   & 0.2823 & 0.2840 & 0.3099 & 0.3101 & 0.3378 & 0.3366 \\
$\Omega_{b}h^2$ & 0.02202 & 0.02201  & 0.02202 & 0.02203  & 0.02202 & 0.02202 & 0.02201 & 0.02202\\
$\Omega_k$      & \multicolumn{2}{c}{---} & 0.063 & 0.057  & \multicolumn{2}{c}{---} & \multicolumn{2}{c}{---} \\
$\omega_0$  & \multicolumn{2}{c}{---} & \multicolumn{2}{c}{---} & \multicolumn{2}{c}{---} & $-0.637$ & $-0.647$  \\
$\omega_a$  & \multicolumn{2}{c}{---} & \multicolumn{2}{c}{---}  & \multicolumn{2}{c}{---}  & $-1.53$ & $-1.49$ \\
\hline
$-2\ln\mathcal{L}_{\text{max}}$  & ~\,12.7190 & 12.7195 & 12.018 & 12.018 & 40.982 & 40.982 & 32.046 & 32.094\\
ncalls & 2,344 & 16,804 & 4,740 & 18,726 & 6,600 & 14,238 & 7,920 & 20,497 \\
\hline
$\Delta$AIC & \multicolumn{2}{c}{---} & 1.298 & 1.298 & \multicolumn{2}{c}{---} & $-4.936$ & $-4.925$\\
$\Delta$BIC & \multicolumn{2}{c}{---} & 1.783 & 1.783 & \multicolumn{2}{c}{---} & $-5.409$ & $-5.398$\\
\bottomrule
\end{tabular}

\caption{Parameter estimation comparison among the PSO algorithm and MCMC, for the different models and dataset combinations. The bottom rows report the minimum chi-squared achieved, or equivalently the $-2\ln\mathcal{L}_{\text{max}}$, the number of likelihood evaluations for each algorithm (ncalls); $\Delta$AIC and $\Delta$BIC values computed relative to flat $\Lambda$CDM.} 
\label{tab:parametros_pso}

\end{table*}

To gain deeper insight into the algorithm’s behavior, we examine the motion of the particles within the parameter space. 
Figures~\ref{fig:Ok_Om}
and~\ref{fig:2dposteriors} offer complementary perspectives on the convergence of the PSO algorithm in the parameter space for the $\Lambda$CDM+$\Omega_k$ model, depicting the locations of particles in the $(\Omega_m, \Omega_k)$ plane at different stages or iterations of the optimization process.
Figure~\ref{fig:Ok_Om} shows how the particles gradually cluster around the best-fit point as the number of iterations increases (from blue to pink, as indicated in the color bar). Once the best-fit values have been identified, we overlay the 68\% and 95\% confidence regions derived from a Fisher matrix analysis (solid red curves); for comparison, we also show the 1 and 2$\sigma$ contour levels obtained with the MCMC algorithm (contours colored purple). The resulting constraints from both methods are qualitatively indistinguishable, while PSO achieves them roughly with fewer likelihood evaluations, as shown in Table \ref{tab:parametros_pso}; both algorithms implemented with the same level of parallelization within the cluster.  
Figure~\ref{fig:2dposteriors} complements this view by presenting the particle distributions iteration by iteration in the same parameter space. The rightmost panel displays the evolution of the cost function in terms of the iteration number, summarizing the efficiency of the optimization. Note that after about 50 iterations, the algorithm has nearly converged, yet it continues to refine the solution until the target accuracy is reached.  
The best-fit parameters for the $\Lambda$CDM+$\Omega_k$ model obtained from DESI BAO data are in agreement with those reported by the DESI DR1 collaboration~\cite{DESI:2024mwx}. 
This level of consistency demonstrates the reliability of the PSO algorithm for constraining cosmological parameters from observational data and shows that, under suitable conditions, PSO can provide results comparable to standard MCMC-based inference pipelines in significantly reduced computational time.

\begin{figure}[h]
\centering
   \includegraphics[trim = 0mm  0mm 0mm 0mm, clip, width=9.cm, height=7.cm]{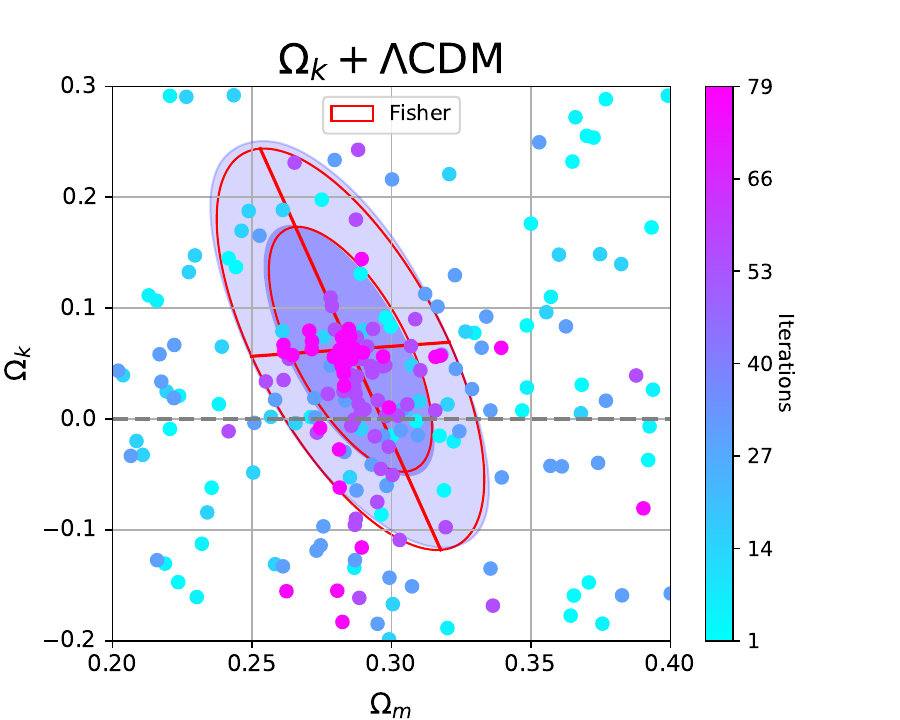}
   \caption{Evolution of $(\Omega_m, \Omega_k)$ points during the PSO optimization for the $\Lambda$CDM+$\Omega_k$ model and using DESI data. The color bar indicates the number of iterations, and the dashed horizontal line represents the flat case. Solid red curves: 1$\sigma$ and 2$\sigma$ confidence regions derived from a Fisher matrix analysis, 
   whereas  contour levels obtained with the MCMC algorithm are purple colored. }
   \label{fig:Ok_Om}
\end{figure}

\begin{figure*}[t]
    \centering
    \makebox[11cm][c]{
\includegraphics[trim = 0mm  0mm 10mm 0mm, clip, width=3.5cm, height=4.4cm]{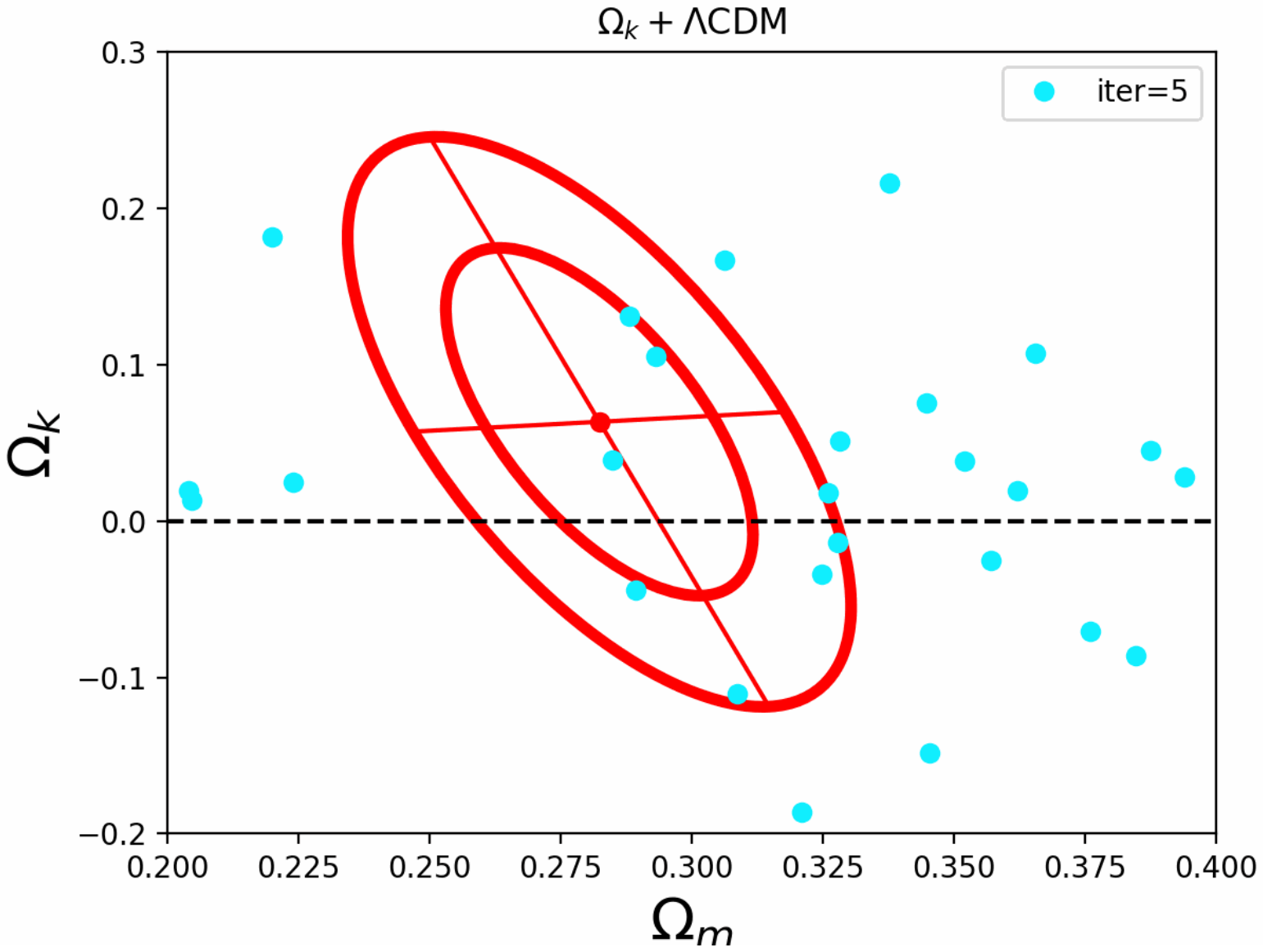}
\includegraphics[trim = 13mm  0mm 10mm 0mm, clip, width=3.5cm, height=4.4cm]{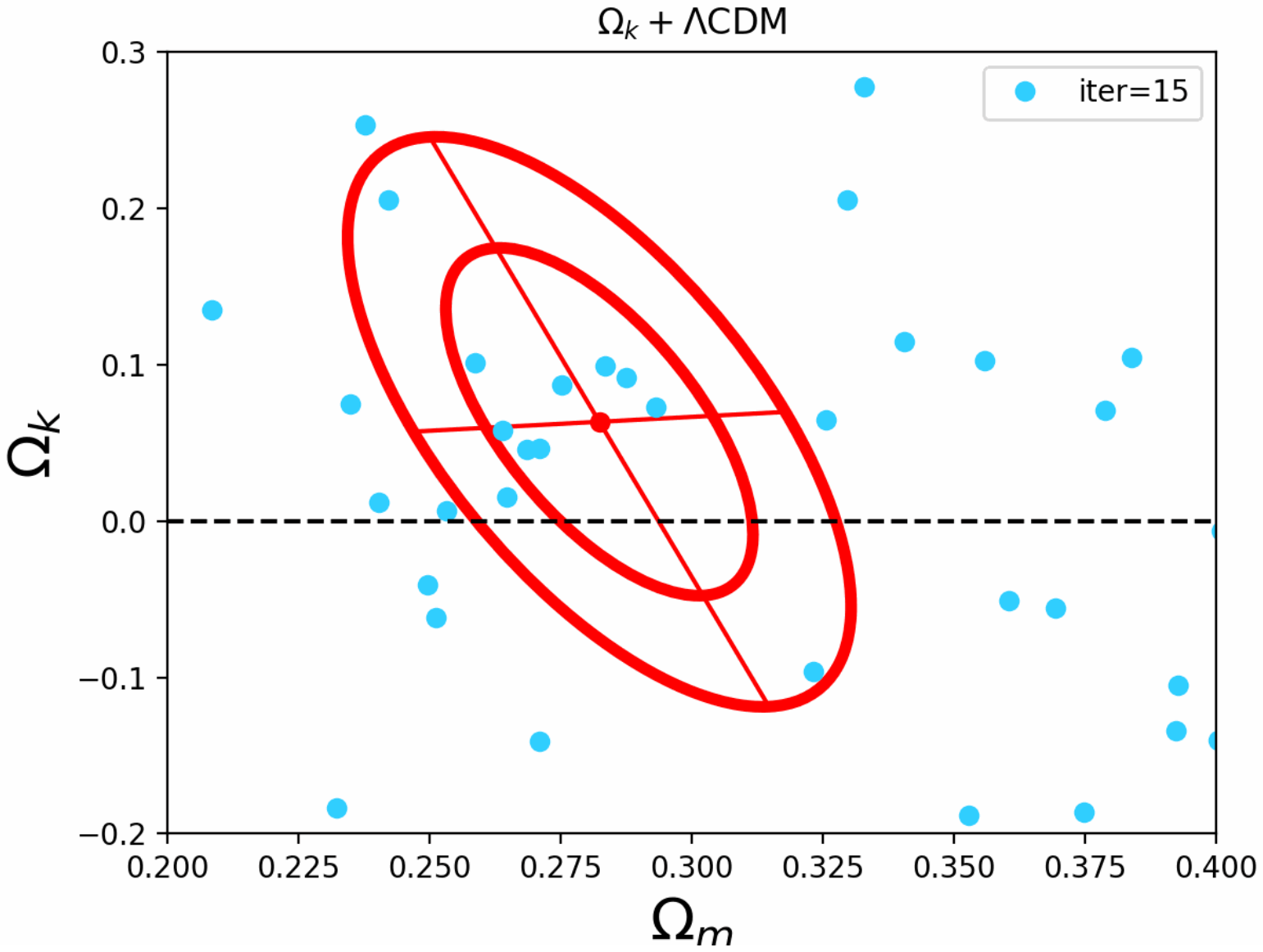}
\includegraphics[trim = 13mm  0mm 10mm 0mm, clip, width=3.5cm, height=4.4cm]{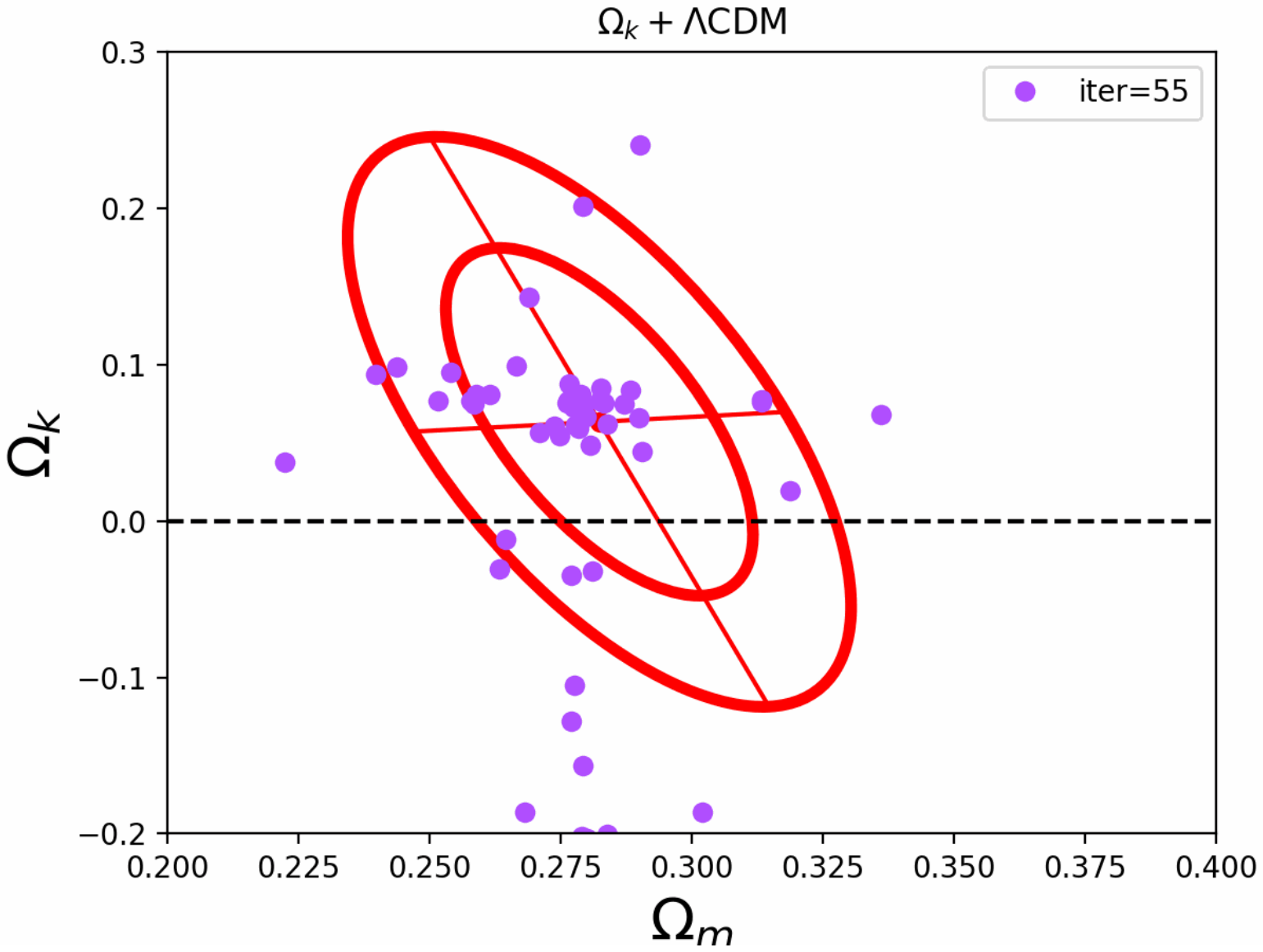}
\includegraphics[trim = 13mm  0mm 10mm 0mm, clip, width=3.5cm, height=4.4cm]{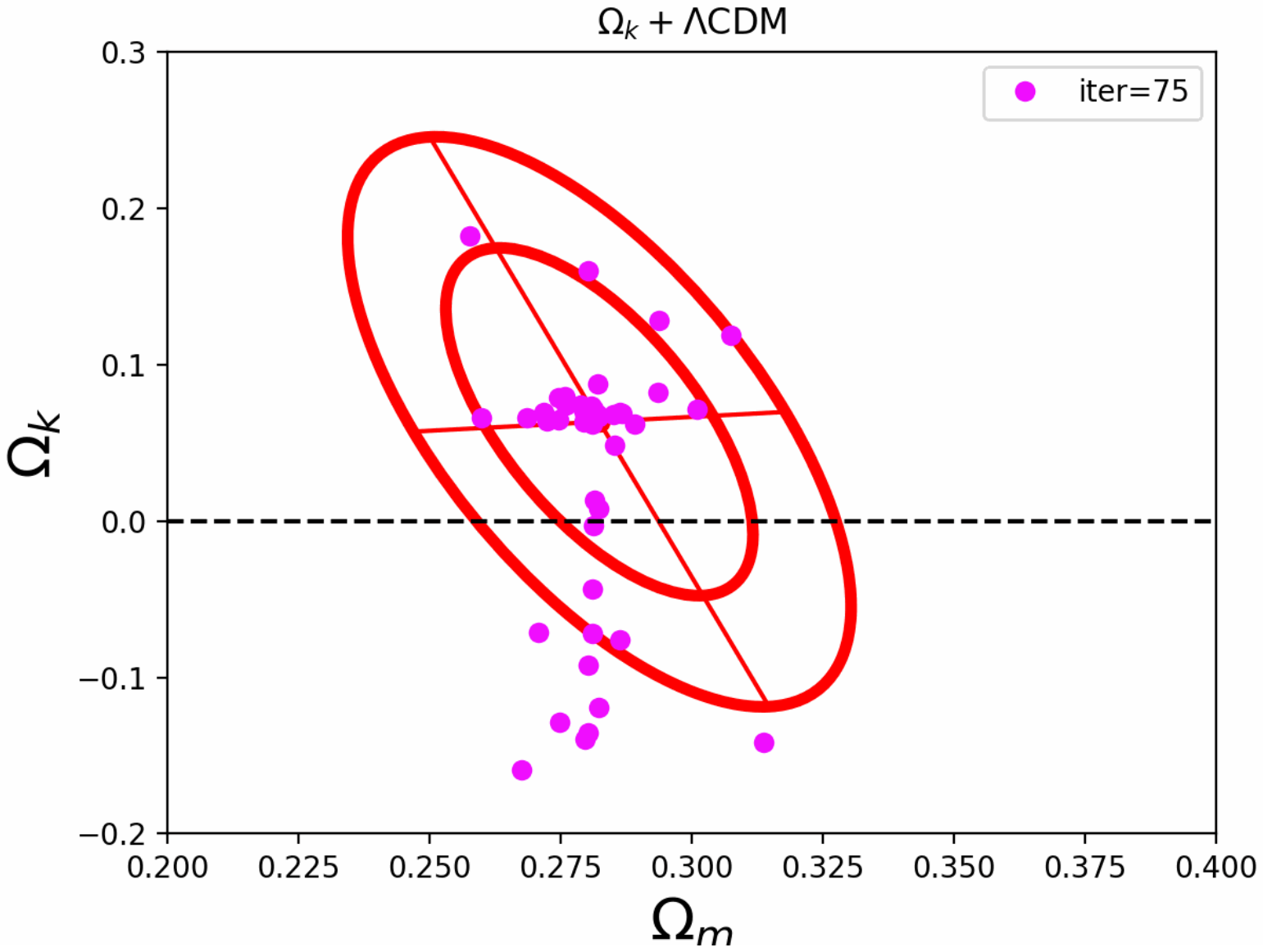}
  \includegraphics[trim = 0mm  0mm 0mm 0mm, clip, width=4.2cm, height=4.5cm]{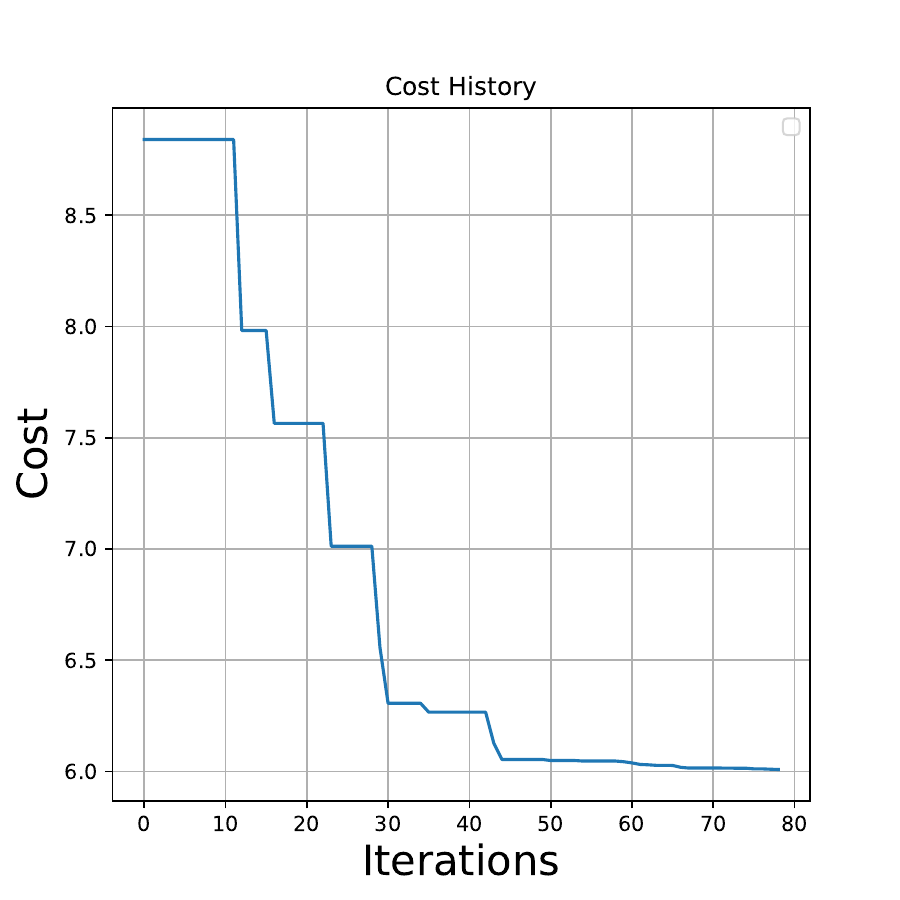}
}

   \caption{Sequential evolution of the particle distribution in the $(\Omega_m, \Omega_k)$ as the iterations of PSO progresses. The right panel displays the cost function value in terms of the number of iterations.}
   \label{fig:2dposteriors}
\end{figure*}

We now focus on the CPL model, whose upper panel in Figure~\ref{fig:wowaCDM} shows how the particles are distributed in the $(\omega_0, \omega_a)$ plane over successive iterations. In this case, by construction, we initialized the entire swarm far outside the target region to show how the particles’ capacity for flexible, interconnected decision-making enabled the swarm to locate a favorable region (around iteration $\sim$18) and subsequently climb toward the peak of the likelihood function. 
The gray dashed lines denote the $\Lambda$CDM boundary, while the 68\% and 95\% confidence contours derived from a Fisher matrix analysis are indicated in red for comparison. As in the previous example, we also overlay the contour levels resulting from the MCMC analysis (purple for 1$\sigma$ and lilac for 2$\sigma$), and the agreement between the contours is almost exact, except for a small deviation in the upper part of the ellipse.  
This representation demonstrates that, as the number of iterations grows (with particle colors evolving from blue to pink), the swarm gradually converges toward the best-fit region, indicated by the cross formed by the red lines. The lower panel supports this conclusion, showing that the fitness function ($\chi^2$) decreases as the iterations progress.

\begin{figure}[h]

   \includegraphics[trim = 0mm  0mm 0mm 0mm, clip, width=8.cm, height=7.cm]{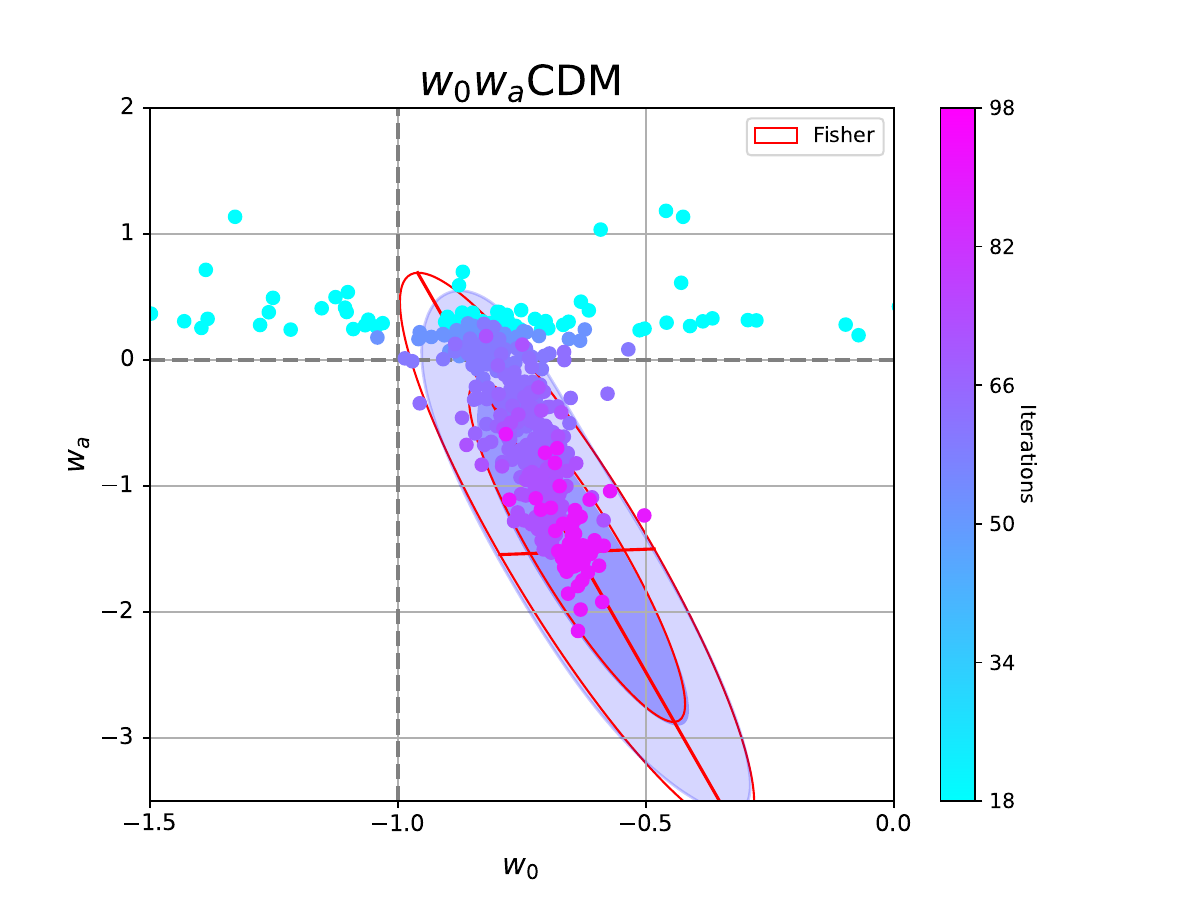}
\includegraphics[trim = 0mm  0mm 0mm 0mm, clip, width=7.cm, height=6.cm]{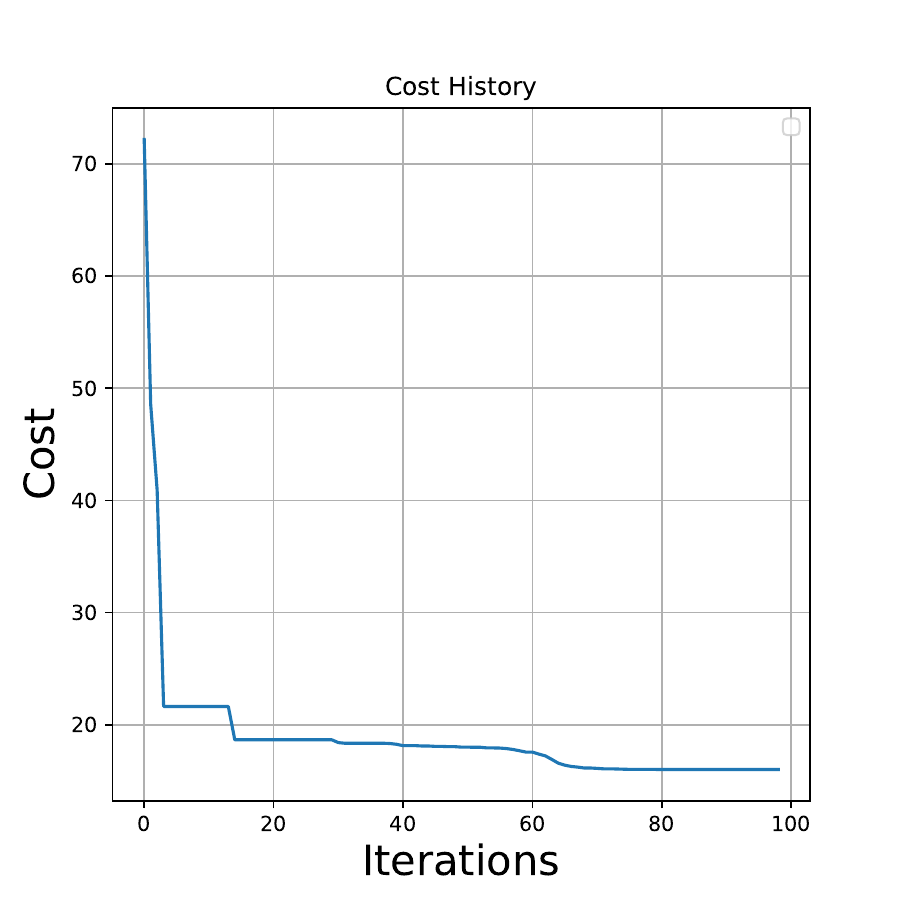}
   \caption{Top panel: Distribution of particles in the $(\omega_0, \omega_a)$ plane during the PSO process for the CPL model, using DESI + Union3 data. 
   The color bar indicates the number of iterations, and the gray dashed lines displays the $\Lambda$CDM limit. Solid red curves: 1$\sigma$ and 2$\sigma$ confidence regions derived from a Fisher matrix analysis, whereas  contour levels obtained with the MCMC algorithm are purple colored.
    Bottom panel: The cost function value versus the iteration number.}
   \label{fig:wowaCDM}
\end{figure}

\section{Conclusions} \label{Sec:conclusions}

In this work, we have presented and assessed the Particle Swarm Optimization algorithm as an efficient tool for estimating free cosmological parameters across different dark energy models. Through its mechanisms of communication and exploration within the swarm, PSO becomes a powerful method for locating global optima in high-dimensional or multimodal functions. 

We investigated the performance of the \textit{Gbest} and \textit{Lbest} variants of PSO in identifying global minima over one- and two-dimensional test-function landscapes. Our findings indicate that both approaches are effective at locating the optimal fitness. Nonetheless, \textit{Gbest} requires only three hyperparameters ($c_{1}, c_{2}$ and $\omega$), whereas \textit{Lbest} needs five to achieve a more thorough exploration, particularly when local minima appear during the search. Motivated by this, we also employed a toy model with synthetic data to test the \textit{Gbest} PSO and found that it is likewise able to recover parameter values. In addition, the paper analyzed the required PSO hyperparameters and the swarm topology needed to ensure efficient communication among particles.

We emphasize that our use of PSO focuses on maximizing the likelihood function and is not meant as a substitute for MCMC methods. Rather, PSO provides complementary benefits, such as faster convergence and accurate best-fit estimates, which can be combined with MCMC approaches. For instance, once PSO has identified the best-fit parameters and the confidence contours have been derived (perhaps via Fisher matrix analysis), these initial values and covariance matrix can be used as informative starting points along with its proposal distribution for MCMC,  thereby accelerating the overall analysis. Such features are especially relevant in cosmology, given the vast amount of current and upcoming data, the high dimensionality of the parameter space, and the complexity of proposed cosmological models.

From the cosmological standpoint, our best-fit estimates agree well with those obtained from MCMC. Notably, when using only DESI data, the flat $\Lambda$CDM model is slightly favored over its $\Omega_k$ extension. However, when a dynamical dark energy equation of state is allowed, the CPL model is preferred over flat $\Lambda$CDM according to both the AIC and BIC, using the combined DESI and Union3 datasets. These results not only reinforce the robustness of the standard cosmological model under current observations but also highlight the potential importance of evolving dark energy scenarios. In this setting, PSO emerges as a powerful and versatile optimization technique, capable of efficiently probing high-dimensional parameter spaces and aiding in cosmological model selection.

\section*{Acknowledgment}
GG-A acknowledge the support of the SECIHTI. J.A.V. acknowledges support from FOSEC SEP-CONACYT Ciencia B\'asica A1-S-21925,  UNAM-DGAPA-PAPIIT IN109126, IN110325 and Cátedra de Investigación Marcos Moshinsky. Special acknowledgments to Ing. Francisco Bustos and Lic. Reyes García who assisted considerably with the High Performance Computing at the ICF-UNAM.
\bibliography{bibliography.bib}

@ARTICLE{slipher1913radial,
       author = {{Slipher}, V.~M.},
        title = "{The radial velocity of the Andromeda Nebula}",
      journal = {Lowell Observatory Bulletin},
         year = 1913,
        month = jan,
       volume = {2},
       number = {8},
        pages = {56-57},
       adsurl = {https://ui.adsabs.harvard.edu/abs/1913LowOB...2...56S}
}

@article{hubble1929relation,
    author = "Hubble, Edwin",
    title = "{A relation between distance and radial velocity among extra-galactic nebulae}",
    doi = "10.1073/pnas.15.3.168",
    journal = "Proc. Nat. Acad. Sci.",
    volume = "15",
    pages = "168--173",
    year = "1929"
}

@article{aghanim2020planck,
    author = "Aghanim, N. and others",
    collaboration = "Planck",
    title = "{Planck 2018 results. VI. Cosmological parameters}",
    eprint = "1807.06209",
    archivePrefix = "arXiv",
    primaryClass = "astro-ph.CO",
    doi = "10.1051/0004-6361/201833910",
    journal = "Astron. Astrophys.",
    volume = "641",
    pages = "A6",
    year = "2020",
    note = "[Erratum: Astron.Astrophys. 652, C4 (2021)]"
}

@article{Medel-Esquivel:2023nov,
    author = "Medel-Esquivel, Ricardo and G{\'o}mez-Vargas, Isidro and S{\'a}nchez, Alejandro A. Morales and Garc{\'\i}a-Salcedo, Ricardo and Alberto V{\'a}zquez, Jos{\'e}",
    title = "{Cosmological Parameter Estimation with Genetic Algorithms}",
    eprint = "2311.05699",
    archivePrefix = "arXiv",
    primaryClass = "astro-ph.CO",
    doi = "10.3390/universe10010011",
    journal = "Universe",
    volume = "10",
    number = "1",
    pages = "11",
    year = "2024"
}

@article{Akarsu:2019hmw,
    author = {Akarsu, {\"O}zg{\"u}r and Barrow, John D. and Escamilla, Luis A. and Vazquez, J. Alberto},
    title = "{Graduated dark energy: Observational hints of a spontaneous sign switch in the cosmological constant}",
    eprint = "1912.08751",
    archivePrefix = "arXiv",
    primaryClass = "astro-ph.CO",
    doi = "10.1103/PhysRevD.101.063528",
    journal = "Phys. Rev. D",
    volume = "101",
    number = "6",
    pages = "063528",
    year = "2020"
}

@article{shi2012comprehensive,
    author = "Shi, Ke and Huang, Yongfeng and Lu, Tan",
    title = "{A comprehensive comparison of cosmological models from latest observational data}",
    eprint = "1207.5875",
    archivePrefix = "arXiv",
    primaryClass = "astro-ph.CO",
    doi = "10.1111/j.1365-2966.2012.21784.x",
    journal = "Mon. Not. Roy. Astron. Soc.",
    volume = "426",
    pages = "2452--2462",
    year = "2012"
}

@article{Chevallier:2000qy,
    author = "Chevallier, Michel and Polarski, David",
    title = "{Accelerating universes with scaling dark matter}",
    eprint = "gr-qc/0009008",
    archivePrefix = "arXiv",
    doi = "10.1142/S0218271801000822",
    journal = "Int. J. Mod. Phys. D",
    volume = "10",
    pages = "213--224",
    year = "2001"
}

@article{Linder:2002et,
    author = "Linder, Eric V.",
    title = "{Exploring the expansion history of the universe}",
    eprint = "astro-ph/0208512",
    archivePrefix = "arXiv",
    doi = "10.1103/PhysRevLett.90.091301",
    journal = "Phys. Rev. Lett.",
    volume = "90",
    pages = "091301",
    year = "2003"
}

@article{scherrer2015mapping,
    author = "Scherrer, Robert J.",
    title = "{Mapping the Chevallier-Polarski-Linder parametrization onto Physical Dark Energy Models}",
    eprint = "1505.05781",
    archivePrefix = "arXiv",
    primaryClass = "astro-ph.CO",
    doi = "10.1103/PhysRevD.92.043001",
    journal = "Phys. Rev. D",
    volume = "92",
    number = "4",
    pages = "043001",
    year = "2015"
}

@article{heavens2009statistical,
    author = "Heavens, Alan",
    title = "{Statistical techniques in cosmology}",
    eprint = "0906.0664",
    archivePrefix = "arXiv",
    journal= " ",
    primaryClass = "astro-ph.CO",
    month = "6",
    year = "2009"
}

@inproceedings{trotta2017bayesian,
    author = "Trotta, Roberto",
    title = "{Bayesian Methods in Cosmology}",
    eprint = "1701.01467",
    booktitle = " ",
    archivePrefix = "arXiv",
    primaryClass = "astro-ph.CO",
    month = "1",
    year = "2017"
}

@article{padilla2021cosmological,
    author = "Padilla, Luis E. and Tellez, Luis O. and Escamilla, Luis A. and Vazquez, Jose Alberto",
    title = "{Cosmological Parameter Inference with Bayesian Statistics}",
    eprint = "1903.11127",
    archivePrefix = "arXiv",
    primaryClass = "astro-ph.CO",
    doi = "10.3390/universe7070213",
    journal = "Universe",
    volume = "7",
    number = "7",
    pages = "213",
    year = "2021"
}

@article{prasad2012cosmological,
    author = "Prasad, Jayanti and Souradeep, Tarun",
    title = "{Cosmological parameter estimation using Particle Swarm Optimization (PSO)}",
    eprint = "1108.5600",
    archivePrefix = "arXiv",
    primaryClass = "astro-ph.CO",
    doi = "10.1103/PhysRevD.85.123008",
    journal = "Phys. Rev. D",
    volume = "85",
    number = "12",
    pages = "123008",
    year = "2012",
    note = "[Erratum: Phys.Rev.D 90, 109903 (2014)]"
}

@article{eisenstein2005detection,
    author = "Eisenstein, Daniel J. and others",
    collaboration = "SDSS",
    title = "{Detection of the Baryon Acoustic Peak in the Large-Scale Correlation Function of SDSS Luminous Red Galaxies}",
    eprint = "astro-ph/0501171",
    archivePrefix = "arXiv",
    reportNumber = "FERMILAB-PUB-05-057-A-CD",
    doi = "10.1086/466512",
    journal = "Astrophys. J.",
    volume = "633",
    pages = "560--574",
    year = "2005"
}

@article{suzuki2012hubble,
    author = "Suzuki, N. and others",
    collaboration = "Supernova Cosmology Project",
    title = "{The Hubble Space Telescope Cluster Supernova Survey: V. Improving the Dark Energy Constraints Above z{\ensuremath{>}}1 and Building an Early-Type-Hosted Supernova Sample}",
    eprint = "1105.3470",
    archivePrefix = "arXiv",
    primaryClass = "astro-ph.CO",
    doi = "10.1088/0004-637X/746/1/85",
    journal = "Astrophys. J.",
    volume = "746",
    pages = "85",
    year = "2012"
}

@article{Amendola:2020qkb,
    author = "Amendola, Luca and G{\'o}mez-Valent, Adri{\`a}",
    title = "{Boosting Monte Carlo sampling with a non-Gaussian fit}",
    eprint = "2007.02615",
    archivePrefix = "arXiv",
    primaryClass = "astro-ph.CO",
    doi = "10.1093/mnras/staa2362",
    journal = "Mon. Not. Roy. Astron. Soc.",
    volume = "498",
    number = "1",
    pages = "181--193",
    year = "2020"
}

@article{darwish2018bio,
  title={Bio-inspired computing: Algorithms review, deep analysis, and the scope of applications},
  author={Darwish, Ashraf},
  journal={Future Computing and Informatics Journal},
  volume={3},
  number={2},
  pages={231--246},
  year={2018},
  publisher={Elsevier}
}

@misc{mratinkovic2019illustrated,
  title={Illustrated Handbook of Particle Swarm Optimisation},
  author={Mratinkovi{\'c}, Aleksandar},
  year={2019},
  publisher={3G E-learning LLC}
}

@book{engelbrecht2007computational,
  title={Computational intelligence: an introduction},
  author={Engelbrecht, Andries P},
  year={2007},
  publisher={John Wiley \& Sons}, 
  address   = {Hoboken, NJ},
  isbn      = {9780470741997},
}

@inproceedings{kennedy1995particle,
    author = "Kennedy, J. and Eberhart, R.",
    title = "{Particle swarm optimization}",
    doi = "10.1109/ICNN.1995.488968",
    year = "1995"
}

@article{parsopoulos2010particle,
  title={Particle swarm optimization and intelligence: advances and applications: advances and applications},
  journal={},
  author={Parsopoulos, Konstantinos E and Vrahatis, Michael N},
  year={2010},
  publisher={IGI global}
}

@inproceedings{shi2001particle,
  title={Particle swarm optimization: developments, applications and resources},
  author={Shi, Yuhui and others},
  booktitle={Proceedings of the 2001 congress on evolutionary computation (IEEE Cat. No. 01TH8546)},
  volume={1},
  pages={81--86},
  year={2001},
  organization={IEEE}
}

@article{mohanty2012particle,
  title={Particle Swarm Optimization and regression analysis--I},
  author={Mohanty, Soumya D},
  journal={Astronomical Review},
  volume={7},
  number={2},
  pages={29--35},
  year={2012},
  publisher={Taylor \& Francis}
}

@article{skokos2005particle,
    author = "Skokos, Charalampos and Parsopoulos, K. E. and Patsis, P. A. and Vrahatis, M. N.",
    title = "{Particle swarm optimization: An Efficient method for tracing periodic orbits in 3-D Galactic potentials}",
    eprint = "astro-ph/0502164",
    archivePrefix = "arXiv",
    doi = "10.1111/j.1365-2966.2005.08892.x",
    journal = "Mon. Not. Roy. Astron. Soc.",
    volume = "359",
    pages = "251--260",
    year = "2005"
}

@book{clerc2010particle,
  title={Particle swarm optimization},
  author={Clerc, Maurice},
  volume={93},
  year={2010},
  publisher={John Wiley \& Sons}
}

@misc{miranda2020pyswarms,
  title={Pyswarms documentation},
  author={Miranda, Lester James V},
  year={2020}
}

@inproceedings{engelbrecht2013particle,
  title={Particle swarm optimization: Global best or local best?},
  author={Engelbrecht, Andries Petrus},
  booktitle={2013 BRICS congress on computational intelligence and 11th Brazilian congress on computational intelligence},
  pages={124--135},
  year={2013},
  organization={IEEE}
}

@article{riess1998observational,
    author = "Riess, Adam G. and others",
    collaboration = "Supernova Search Team",
    title = "{Observational evidence from supernovae for an accelerating universe and a cosmological constant}",
    eprint = "astro-ph/9805201",
    archivePrefix = "arXiv",
    doi = "10.1086/300499",
    journal = "Astron. J.",
    volume = "116",
    pages = "1009--1038",
    year = "1998"
}

@article{SDSS:2003eyi,
    author = "Tegmark, Max and others",
    collaboration = "SDSS",
    title = "{Cosmological parameters from SDSS and WMAP}",
    eprint = "astro-ph/0310723",
    archivePrefix = "arXiv",
    reportNumber = "FERMILAB-PUB-03-435-A",
    doi = "10.1103/PhysRevD.69.103501",
    journal = "Phys. Rev. D",
    volume = "69",
    pages = "103501",
    year = "2004"
}

@article{perlmutter1999measurements,
    author = "Perlmutter, S. and others",
    collaboration = "Supernova Cosmology Project",
    title = "{Measurements of $\Omega$ and $\Lambda$ from 42 High Redshift Supernovae}",
    eprint = "astro-ph/9812133",
    archivePrefix = "arXiv",
    reportNumber = "LBNL-41801, LBL-41801",
    doi = "10.1086/307221",
    journal = "Astrophys. J.",
    volume = "517",
    pages = "565--586",
    year = "1999"
}

@article{copeland2006dynamics,
    author = "Copeland, Edmund J. and Sami, M. and Tsujikawa, Shinji",
    title = "{Dynamics of dark energy}",
    eprint = "hep-th/0603057",
    archivePrefix = "arXiv",
    doi = "10.1142/S021827180600942X",
    journal = "Int. J. Mod. Phys. D",
    volume = "15",
    pages = "1753--1936",
    year = "2006"
}

@article{Bull:2015stt,
    author = "Bull, Philip and others",
    title = "{Beyond $\Lambda$CDM: Problems, solutions, and the road ahead}",
    eprint = "1512.05356",
    archivePrefix = "arXiv",
    primaryClass = "astro-ph.CO",
    doi = "10.1016/j.dark.2016.02.001",
    journal = "Phys. Dark Univ.",
    volume = "12",
    pages = "56--99",
    year = "2016"
}

@article{vazquez2008materia,
  title={La materia oscura del universo: retos y perspectivas},
  author={V{\'a}zquez-Gonz{\'a}lez, A and Matos, T},
  journal={Revista mexicana de f{\'\i}sica E},
  volume={54},
  number={2},
  pages={193--202},
  year={2008},
  publisher={Sociedad Mexicana de F{\'\i}sica}
}

@article{DESI:2025zgx,
    author = "Abdul Karim, M. and others",
    collaboration = "DESI",
    journal="",
    title = "{DESI DR2 Results II: Measurements of Baryon Acoustic Oscillations and Cosmological Constraints}",
    eprint = "2503.14738",
    archivePrefix = "arXiv",
    primaryClass = "astro-ph.CO",
    reportNumber = "FERMILAB-PUB-25-0169-PPD",
    month = "3",
    year = "2025"
}

@article{Riess:2021jrx,
    author = "Riess, Adam G. and others",
    title = "{A Comprehensive Measurement of the Local Value of the Hubble Constant with 1 km s$^{-1}$ Mpc$^{-1}$ Uncertainty from the Hubble Space Telescope and the SH0ES Team}",
    eprint = "2112.04510",
    archivePrefix = "arXiv",
    primaryClass = "astro-ph.CO",
    doi = "10.3847/2041-8213/ac5c5b",
    journal = "Astrophys. J. Lett.",
    volume = "934",
    number = "1",
    pages = "L7",
    year = "2022"
}

@article{poli2008analysis,
  title={Analysis of the publications on the applications of particle swarm optimisation},
  author={Poli, Riccardo},
  journal={Journal of Artificial Evolution and Applications},
  volume={2008},
  number={1},
  pages={685175},
  year={2008},
  publisher={Wiley Online Library}
}

@article{abido2002optimal,
  title={Optimal power flow using particle swarm optimization},
  author={Abido, Mohammad A},
  journal={International Journal of Electrical Power \& Energy Systems},
  volume={24},
  number={7},
  pages={563--571},
  year={2002},
  publisher={Elsevier}
}

@phdthesis{wang2015first,
    author = "Wang, Yan",
    title = "{First-stage LISA data processing and gravitational wave data analysis}: {Ultraprecise inter-satellite laser ranging, clock synchronization and novel gravitational wave data analysis algorithms}",
    doi = "10.1007/978-3-319-26389-2",
    school = "Hannover, Max Planck Inst. Grav.",
    year = "2014"
}

@article{wang2018particle,
  title={Particle swarm optimization algorithm: an overview},
  author={Wang, Dongshu and Tan, Dapei and Liu, Lei},
  journal={Soft computing},
  volume={22},
  number={2},
  pages={387--408},
  year={2018},
  publisher={Springer}
}

@inproceedings{shi1999empirical,
  title={Empirical study of particle swarm optimization},
  author={Shi, Yuhui and Eberhart, Russell C},
  booktitle={Proceedings of the 1999 congress on evolutionary computation-CEC99 (Cat. No. 99TH8406)},
  volume={3},
  pages={1945--1950},
  year={1999},
  organization={IEEE}
}

@article{arun2017dark,
    author = "Arun, Kenath and Gudennavar, S. B. and Sivaram, C.",
    title = "{Dark matter, dark energy, and alternate models: A review}",
    eprint = "1704.06155",
    archivePrefix = "arXiv",
    primaryClass = "physics.gen-ph",
    doi = "10.1016/j.asr.2017.03.043",
    journal = "Adv. Space Res.",
    volume = "60",
    pages = "166--186",
    year = "2017"
}

@article{weinberg1989cosmological,
    author = "Weinberg, Steven",
    editor = "Hsu, Jong-Ping and Fine, D.",
    title = "{The Cosmological Constant Problem}",
    reportNumber = "UTTG-12-88",
    doi = "10.1103/RevModPhys.61.1",
    journal = "Rev. Mod. Phys.",
    volume = "61",
    pages = "1--23",
    year = "1989"
}

@article{Akaike:1974vps,
    author = "Akaike, H.",
    title = "{A new look at the statistical model identification}",
    doi = "10.1109/TAC.1974.1100705",
    journal = "IEEE Trans. Automatic Control",
    volume = "19",
    number = "6",
    pages = "716--723",
    year = "1974"
}

@article{Schwarz:1978tpv,
    author = "Schwarz, Gideon",
    title = "{Estimating the Dimension of a Model}",
    journal = "Annals Statist.",
    volume = "6",
    pages = "461--464",
    year = "1978"
}

@article{verde2010statistical,
    author = "Verde, Licia",
    title = "{Statistical methods in cosmology}",
    eprint = "0911.3105",
    archivePrefix = "arXiv",
    primaryClass = "astro-ph.CO",
    doi = "10.1007/978-3-642-10598-2_4",
    journal = "Lect. Notes Phys.",
    volume = "800",
    pages = "147--177",
    year = "2010"
}

@article{CosmoVerseNetwork:2025alb,
    author = "Di Valentino, Eleonora and others",
    collaboration = "CosmoVerse Network",
    title = "{The CosmoVerse White Paper: Addressing observational tensions in cosmology with systematics and fundamental physics}",
    eprint = "2504.01669",
    archivePrefix = "arXiv",
    primaryClass = "astro-ph.CO",
    doi = "10.1016/j.dark.2025.101965",
    journal = "Phys. Dark Univ.",
    volume = "49",
    pages = "101965",
    year = "2025"
}

@article{Gomez-Vargas:2024izm,
    author = "G{\'o}mez-Vargas, Isidro and V{\'a}zquez, J. Alberto",
    title = "{Deep learning and genetic algorithms for cosmological Bayesian inference speed-up}",
    eprint = "2405.03293",
    archivePrefix = "arXiv",
    primaryClass = "astro-ph.IM",
    doi = "10.1103/PhysRevD.110.083518",
    journal = "Phys. Rev. D",
    volume = "110",
    number = "8",
    pages = "083518",
    year = "2024"
}

@article{Gomez-Vargas:2022bsm,
    author = "G{\'o}mez-Vargas, Isidro and Andrade, Joshua Briones and V{\'a}zquez, J. Alberto",
    title = "{Neural networks optimized by genetic algorithms in cosmology}",
    eprint = "2209.02685",
    archivePrefix = "arXiv",
    primaryClass = "astro-ph.IM",
    doi = "10.1103/PhysRevD.107.043509",
    journal = "Phys. Rev. D",
    volume = "107",
    number = "4",
    pages = "043509",
    year = "2023"
}

@article{Vazquez:2020ani,
    author = "V{\'a}zquez, J. Alberto and Tamayo, David and Sen, Anjan A. and Quiros, Israel",
    title = "{Bayesian model selection on scalar $\epsilon$-field dark energy}",
    eprint = "2009.01904",
    archivePrefix = "arXiv",
    primaryClass = "gr-qc",
    doi = "10.1103/PhysRevD.103.043506",
    journal = "Phys. Rev. D",
    volume = "103",
    number = "4",
    pages = "043506",
    year = "2021"
}

@article{Escamilla:2024fzq,
    author = "Escamilla, Luis A. and Pan, Supriya and Di Valentino, Eleonora and Paliathanasis, Andronikos and V{\'a}zquez, Jos{\'e} Alberto and Yang, Weiqiang",
    title = "{Testing an oscillatory behavior of dark energy}",
    eprint = "2404.00181",
    archivePrefix = "arXiv",
    primaryClass = "astro-ph.CO",
    doi = "10.1103/PhysRevD.111.023531",
    journal = "Phys. Rev. D",
    volume = "111",
    number = "2",
    pages = "023531",
    year = "2025"
}

@misc{simplemc_github,
  author  = {Vazquez, J. A. and Gomez-Vargas, I. and Slosar, A.},
  title   = {SimpleMC: a simple MCMC code for cosmological parameter estimation},
  url     = {https://github.com/ja-vazquez/SimpleMC},
  note    = {GitHub repository, accessed 2025-12-16}
}

@article{Speagle:2019ivv,
    author = "Speagle, Joshua S.",
    title = "{dynesty: a dynamic nested sampling package for estimating Bayesian posteriors and evidences}",
    eprint = "1904.02180",
    archivePrefix = "arXiv",
    primaryClass = "astro-ph.IM",
    doi = "10.1093/mnras/staa278",
    journal = "Mon. Not. Roy. Astron. Soc.",
    volume = "493",
    number = "3",
    pages = "3132--3158",
    year = "2020"
}

@article{BOSS:2014hhw,
    author = "Aubourg, \'Eric and others",
    collaboration = "BOSS",
    title = "{Cosmological implications of baryon acoustic oscillation measurements}",
    eprint = "1411.1074",
    archivePrefix = "arXiv",
    primaryClass = "astro-ph.CO",
    doi = "10.1103/PhysRevD.92.123516",
    journal = "Phys. Rev. D",
    volume = "92",
    number = "12",
    pages = "123516",
    year = "2015"
}

@article{Rubin:2023ovl,
    author = "Rubin, David and others",
    title = "{Union Through UNITY: Cosmology with 2,000 SNe Using a Unified Bayesian Framework}",
    journal="",
    eprint = "2311.12098",
    archivePrefix = "arXiv",
    primaryClass = "astro-ph.CO",
    month = "11",
    year = "2023"
}

@article{DESI:2024mwx,
    author = "Adame, A. G. and others",
    collaboration = "DESI",
    title = "{DESI 2024 VI: Cosmological Constraints from the Measurements of Baryon Acoustic Oscillations}",
    journal= " ",
    eprint = "2404.03002",
    archivePrefix = "arXiv",
    primaryClass = "astro-ph.CO",
    reportNumber = "FERMILAB-PUB-24-0154-PPD",
    month = "4",
    year = "2024"
}

@article{Zuntz:2014csq,
    author = "Zuntz, Joe and Paterno, Marc and Jennings, Elise and Rudd, Douglas and Manzotti, Alessandro and Dodelson, Scott and Bridle, Sarah and Sehrish, Saba and Kowalkowski, James",
    title = "{CosmoSIS: modular cosmological parameter estimation}",
    eprint = "1409.3409",
    archivePrefix = "arXiv",
    primaryClass = "astro-ph.CO",
    reportNumber = "FERMILAB-PUB-14-408-A",
    doi = "10.1016/j.ascom.2015.05.005",
    journal = "Astron. Comput.",
    volume = "12",
    pages = "45--59",
    year = "2015"
}

@article{Torrado:2020dgo,
    author = "Torrado, Jesus and Lewis, Antony",
    title = "{Cobaya: Code for Bayesian Analysis of hierarchical physical models}",
    eprint = "2005.05290",
    archivePrefix = "arXiv",
    primaryClass = "astro-ph.IM",
    reportNumber = "TTK-20-15",
    doi = "10.1088/1475-7516/2021/05/057",
    journal = "JCAP",
    volume = "05",
    pages = "057",
    year = "2021"
}

@article{WMAP:2010qai,
    author = "Komatsu, E. and others",
    collaboration = "WMAP",
    title = "{Seven-Year Wilkinson Microwave Anisotropy Probe (WMAP) Observations: Cosmological Interpretation}",
    eprint = "1001.4538",
    archivePrefix = "arXiv",
    primaryClass = "astro-ph.CO",
    doi = "10.1088/0067-0049/192/2/18",
    journal = "Astrophys. J. Suppl.",
    volume = "192",
    pages = "18",
    year = "2011"
}

@article{Toni_2008,
   title={Approximate Bayesian computation scheme for parameter inference and model selection in dynamical systems},
   volume={6},
   ISSN={1742-5662},
   url={http://dx.doi.org/10.1098/rsif.2008.0172},
   DOI={10.1098/rsif.2008.0172},
   number={31},
   journal={Journal of The Royal Society Interface},
   publisher={The Royal Society},
   author={Toni, Tina and Welch, David and Strelkowa, Natalja and Ipsen, Andreas and Stumpf, Michael P.H},
   year={2008},
   month=July, pages={187–202} }

@misc{toni2010simulationbasedmodelselectiondynamical,
      title={Simulation-based model selection for dynamical systems in systems and population biology}, 
      author={Tina Toni and Michael P. H. Stumpf},
      year={2010},
      eprint={0911.1705},
      archivePrefix={arXiv},
      primaryClass={q-bio.QM},
      url={https://arxiv.org/abs/0911.1705}, 
}

@article{Bernardo:2022pyz,
    author = "Bernardo, Reginald Christian and Grand{\'o}n, Daniela and Levi Said, Jackson and C{\'a}rdenas, V{\'\i}ctor H.",
    title = "{Dark energy by natural evolution: Constraining dark energy using Approximate Bayesian Computation}",
    eprint = "2211.05482",
    archivePrefix = "arXiv",
    primaryClass = "astro-ph.CO",
    doi = "10.1016/j.dark.2023.101213",
    journal = "Phys. Dark Univ.",
    volume = "40",
    pages = "101213",
    year = "2023"
}

@article{Bernardo:2022ggl,
    author = "Bernardo, Reginald Christian and Lee, You-Ru",
    title = "{Hubble constant by natural selection: Evolution chips in the Hubble tension}",
    eprint = "2212.02203",
    archivePrefix = "arXiv",
    primaryClass = "astro-ph.CO",
    doi = "10.1016/j.ascom.2023.100740",
    journal = "Astron. Comput.",
    volume = "44",
    pages = "100740",
    year = "2023"
}

@article{kennedy2001morgan,
  title={The Morgan Kaufmann series in evolutionary computation},
  author={Kennedy, J and Eberhart, RC and Shi, Yuhui},
  year={2001},
  journal={},
  publisher={Morgan Kaufmann Publishers}
}

@inproceedings{kennedy2002population,
  title={Population structure and particle swarm performance},
  author={Kennedy, James and Mendes, Rui},
  booktitle={Proceedings of the 2002 Congress on Evolutionary Computation. CEC'02 (Cat. No. 02TH8600)},
  volume={2},
  pages={1671--1676},
  year={2002},
  organization={IEEE}
}

@article{kennedy2006neighborhood,
  title={Neighborhood topologies in fully informed and best-of-neighborhood particle swarms},
  author={Kennedy, James and Mendes, Rui},
  journal={IEEE Transactions on Systems, Man, and Cybernetics, Part C (Applications and Reviews)},
  volume={36},
  number={4},
  pages={515--519},
  year={2006},
  publisher={IEEE}
}

@inproceedings{shi1998parameter,
  title={Parameter selection in particle swarm optimization},
  author={Shi, Yuhui and Eberhart, Russell C},
  booktitle={Evolutionary Programming VII: 7th International Conference, EP98 San Diego, California, USA, March 25--27, 1998 Proceedings 7},
  pages={591--600},
  year={1998},
  organization={Springer}
}

@article{lu2015generalized,
  title={Generalized radial basis function neural network based on an improved dynamic particle swarm optimization and AdaBoost algorithm},
  author={Lu, Jinna and Hu, Hongping and Bai, Yanping},
  journal={Neurocomputing},
  volume={152},
  pages={305--315},
  year={2015},
  publisher={Elsevier}
}

@article{Pantazis:2016nky,
    author = "Pantazis, G. and Nesseris, S. and Perivolaropoulos, L.",
    title = "{Comparison of thawing and freezing dark energy parametrizations}",
    eprint = "1603.02164",
    archivePrefix = "arXiv",
    primaryClass = "astro-ph.CO",
    reportNumber = "IFT-UAM-CSIC-16-023",
    doi = "10.1103/PhysRevD.93.103503",
    journal = "Phys. Rev. D",
    volume = "93",
    number = "10",
    pages = "103503",
    year = "2016"
}

@article{verde2014expansion,
    author = "Verde, Licia and Protopapas, Pavlos and Jimenez, Raul",
    title = "{The expansion rate of the intermediate Universe in light of Planck}",
    eprint = "1403.2181",
    archivePrefix = "arXiv",
    primaryClass = "astro-ph.CO",
    doi = "10.1016/j.dark.2014.09.003",
    journal = "Phys. Dark Univ.",
    volume = "5-6",
    pages = "307--314",
    year = "2014"
}

@article{lewis2002cosmological,
    author = "Lewis, Antony and Bridle, Sarah",
    title = "{Cosmological parameters from CMB and other data: A Monte Carlo approach}",
    eprint = "astro-ph/0205436",
    archivePrefix = "arXiv",
    doi = "10.1103/PhysRevD.66.103511",
    journal = "Phys. Rev. D",
    volume = "66",
    pages = "103511",
    year = "2002"
}

@book{gilks1995markov,
  title={Markov chain Monte Carlo in practice},
  author={Gilks, Walter R and Richardson, Sylvia and Spiegelhalter, David},
  year={1995},
  publisher={CRC press}
}

@inproceedings{reynolds1987flocks,
  title={Flocks, herds and schools: A distributed behavioral model},
  author={Reynolds, Craig W},
  booktitle={Proceedings of the 14th annual conference on Computer graphics and interactive techniques},
  pages={25--34},
  year={1987}
}

@article{freitas2020particle,
  title={Particle swarm optimisation: a historical review up to the current developments},
  author={Freitas, Diogo and Lopes, Luiz Guerreiro and Morgado-Dias, Fernando},
  journal={Entropy},
  volume={22},
  number={3},
  pages={362},
  year={2020},
  publisher={MDPI}
}

@inproceedings{shi1998modified,
  title={A modified particle swarm optimizer},
  author={Shi, Yuhui and Eberhart, Russell},
  booktitle={1998 IEEE international conference on evolutionary computation proceedings. IEEE world congress on computational intelligence (Cat. No. 98TH8360)},
  pages={69--73},
  year={1998},
  organization={Ieee}
}

@article{zhan2009adaptive,
  title={Adaptive particle swarm optimization},
  author={Zhan, Zhi-Hui and Zhang, Jun and Li, Yun and Chung, Henry Shu-Hung},
  journal={IEEE Transactions on Systems, Man, and Cybernetics, Part B (Cybernetics)},
  volume={39},
  number={6},
  pages={1362--1381},
  year={2009},
  publisher={IEEE}
}

@article{van2000cooperative,
  title={Cooperative learning in neural networks using particle swarm optimizers},
  author={Van den Bergh, Frans and Engelbrecht, Andries P},
  journal={South African Computer Journal},
  volume={2000},
  number={26},
  pages={84--90},
  year={2000},
  publisher={South African Computer Society (SAICSIT)}
}

@article{sengupta2018particle,
  title={Particle Swarm Optimization: A survey of historical and recent developments with hybridization perspectives},
  author={Sengupta, Saptarshi and Basak, Sanchita and Peters, Richard Alan},
  journal={Machine learning and knowledge extraction},
  volume={1},
  number={1},
  pages={157--191},
  year={2018},
  publisher={MDPI}
}

@inproceedings{engelbrecht2012particle,
  title={Particle swarm optimization: Velocity initialization},
  author={Engelbrecht, Andries},
  booktitle={2012 IEEE congress on evolutionary computation},
  pages={1--8},
  year={2012},
  organization={IEEE}
}

@inproceedings{helwig2008theoretical,
  title={Theoretical analysis of initial particle swarm behavior},
  author={Helwig, Sabine and Wanka, Rolf},
  booktitle={International conference on parallel problem solving from nature},
  pages={889--898},
  year={2008},
  organization={Springer}
}

\end{document}